\documentclass[numberedappendix]{emulateapj}

\begin{document}
\shorttitle{Diffusion of magnetic field via reconnection}
\shortauthors{Santos-Lima et al.}
\title{Diffusion of magnetic field and removal of magnetic flux from clouds via turbulent reconnection}

\author{R. Santos-Lima\altaffilmark{1,2}, A. Lazarian\altaffilmark{2}, E. M. de Gouveia Dal Pino\altaffilmark{1}, J. Cho\altaffilmark{2,3}}
\altaffiltext{1}{Instituto de Astronomia, Geof\'isica e Ci\^encias Atmosf\'ericas, Universidade de S\~ao Paulo, R. do Mat\~ao, 1226, S\~ao Paulo, SP 05508-090, Brazil}
\altaffiltext{2}{Department of Astronomy, University of Wisconsin, Madison, WI 53706, USA}
\altaffiltext{3}{Department of Astronomy and Space Science, Chungnam National University, Daejeon, Korea}

\begin{abstract}
The diffusion of astrophysical magnetic fields in conducting fluids in the presence of turbulence depends on whether magnetic fields can change their topology via reconnection in highly conducting media. Recent progress in understanding fast magnetic reconnection in the presence of turbulence is reassuring that the magnetic field behavior in computer simulations and turbulent astrophysical environments is similar, as far as magnetic reconnection is concerned. This makes it meaningful to perform MHD simulations of turbulent flows in order to understand the diffusion of magnetic field in astrophysical environments. Our studies of magnetic field diffusion in turbulent medium reveal interesting new phenomena. First of all, our three-dimensional MHD simulations initiated with anti-correlating magnetic field and gaseous density exhibit at later times a de-correlation of the magnetic field and density, which corresponds well to the observations of the interstellar media. While earlier studies stressed the role of either ambipolar diffusion or time-dependent turbulent fluctuations for de-correlating magnetic field and density, we get the effect of {\it permanent} de-correlation with one fluid code, i.e. without invoking ambipolar diffusion. In addition, in the presence of gravity and turbulence, our three-dimensional simulations show the decrease of the magnetic flux-to-mass ratio as the gaseous density at the center of the gravitational potential increases. We observe this effect both in the situations when we start with equilibrium distributions of gas and magnetic field and when we follow the evolution of collapsing dynamically unstable configurations. Thus the process of turbulent magnetic field removal should be applicable both to quasi-static subcritical molecular clouds and cores and violently collapsing supercritical entities. The increase of the gravitational potential as well as the magnetization of the gas increases the segregation of the mass and magnetic flux in the saturated final state of the simulations, supporting the notion that the reconnection-enabled diffusivity relaxes the magnetic field + gas system in the gravitational field to its minimal energy state. This effect is expected to play an important role in star formation, from its initial stages of concentrating interstellar gas to the final stages of the accretion to the forming protostar. In addition, we benchmark our codes by studying the heat transfer in magnetized compressible fluids and confirm the high rates of turbulent advection of heat obtained in an earlier study. 

\end{abstract}

\keywords{diffusion --- ISM: magnetic fields --- magnetohydrodynamics (MHD) --- star: formation --- turbulence }

\section{Introduction}

Astrophysical flows are known to be turbulent and magnetized. The specific role played by MHD turbulence in different branches of astrophysics is still highly debated, but it is generally regarded as important. In particular, for the interstellar medium (ISM) and star formation, the role of turbulence has been discussed in many reviews (see \citealt{elmegreen2004,mckee2007}). The opinion on the role of magnetic field in these environments vary from magnetic field being regarded as absolutely dominant in the processes (see \citealt{tassis2005,galli2006}) to moderately important, as in super-Alfv\'enic models of star formation (see \citealt{padoan2004}).

The vital question that frequently permeates these debates is the diffusion of the magnetic field in astrophysical fluids. The conductivity of most of the astrophysical fluids is high enough to make the Ohmic diffusion negligible on the scales involved, which means that the ``frozen-in'' approximation is a good one for many astrophysical environments. However, without considering diffusive mechanisms that can violate the flux freezing, one faces problems attempting to explain many observational facts. For example, simple estimates show that if all the magnetic flux is brought together with the material that collapses to form a star in molecular clouds, then the magnetic field in a proto-star should be several orders of magnitude higher than the one observed in T-Tauri stars (this is the ``magnetic flux problem'', see \citealt{galli2006} and references therein, for example).

To address the problem of the magnetic field diffusion both in the partially ionized ISM and in molecular clouds, researchers usually appeal to the ambipolar diffusion concept (see \citealt{mestel1956,shu1983}). The idea of the ambipolar diffusion is very simple and may be easily exemplified in the case of gas collapsing to form a protostar. As the magnetic field is acting on charged particles only, it does not directly affect neutrals. Neutrals move under the gravitational pull but are scattered by collisions with ions and charged dust grains which are coupled with the magnetic field. The resulting flow dominated by the neutrals will be unable to drag the magnetic field lines and these will diffuse away through the infalling matter. This process of ambipolar diffusion becomes faster as the ionization ratio decreases and therefore, becomes more important in poorly ionized cloud cores.

\citet{shu2006} have explored the accretion phase in low-mass star formation and concluded that there should exist  an effective diffusivity about 4 orders of magnitude larger than Ohmic diffusivity in order to an efficient magnetic flux transport to occur.  They have argued that ambipolar diffusion could work, but only under  rather special circumstances like, for instance, considering particular dust grain sizes. In other words, at the moment it is unclear if  ambipolar diffusion is  really high enough to solve the magnetic flux transport problem in collapsing flows.

Does magnetic field remain absolutely frozen-in within highly ionized astrophysical fluids? The answer to this question affects the description of numerous essential processes in the interstellar and intergalactic gas.

Magnetic reconnection was appealed in \citet{lazarian2005} as a way of removing magnetic flux from gravitating clouds, e.g. from star-forming clouds. That work referred to the reconnection model of \citet{lazarian1999} and \citet{lazarian2004} for the justification of the concept of fast magnetic reconnection in the presence of turbulence. The advantage of the scheme proposed by \citet{lazarian2005} was that robust removal of magnetic flux can be accomplished both in partially and fully ionized plasma, with only marginal dependence on the ionization state of the gas\footnote{The rates were predicted to depend on the reconnection rate, which according to \citet{lazarian2004} very weakly depends on the ionization degree of the gas.}. The concept of ``reconnection diffusion'' introduced 
in \citet{lazarian2005} is relevant to our understanding of many basic astrophysical processes. In particular, it
suggests that the classical textbook description of molecular clouds supported both by hourglass magnetic field
and turbulence is not self-consistent. Indeed, turbulence is expected to induce ``reconnection diffusion'' which
should enable fast magnetic field removal from the cloud.  However, in the absence of numerical confirmation of the fast reconnection, the scheme of magnetic flux removal through ``reconnection diffusion'' as opposed to ambipolar diffusion stayed somewhat speculative.

Fortunately, it has been recently shown numerically (see \citealt{kowal2009}) 
that three-dimensional magnetic reconnection in turbulent fluid follows the predictions of 
the reconnection model of \citet{lazarian1999} and therefore, is fast. This naturally increased the interest to
the ``reconnection diffusion'' (see \citealt{lazarian2009}). 
Motivated by this fact, here we perform simulations aiming to gain understanding of the diffusion of magnetic field induced by turbulence. As \citet{kowal2009} tested magnetic reconnection in fully ionized gas in the present paper we shall focus our efforts on  one fluid MHD simulations.

We shall compare ``reconnection diffusion'' with the ambipolar diffusion and discuss the effects of ambipolar diffusion qualitatively focusing on the  comparison of our results on ``reconnection diffusion'' with those on ``ambipolar turbulent diffusion'' described in \citet{heitsch2004}. The latter study  reported the enhancement of ambipolar diffusion in the presence of turbulence, which raised the question of how important is the simultaneous action of turbulence and ambipolar diffusion and whether turbulence alone, i.e. without any effect from ambipolar diffusion, can equally well induce de-correlation of magnetic field and density.

Our work on ``reconnection diffusion'' should  also be distinguished from the research on the de-correlation of magnetic field and density within compressible turbulent fluctuations. \citet{cho_lazarian2002,cho_lazarian2003} performed three-dimensional MHD simulations and reported the existence of separate turbulent cascades of Alfv\'en and fast modes in strongly driven turbulence as well as a cascade of slow modes driven by Alfv\'enic cascade. Slow modes in magnetically dominated plasma are associated with density perturbations with marginal perturbation of magnetic fields, while the same is true for fast modes in weakly magnetized or high beta plasmas. Naturally, these two modes de-correlate magnetic fields and density on the crossing time of the wave. This was the effect studied in more detail in one-dimensional both analytically and numerically by \citet{passot2003}, who stressed that the enhancements of magnetic field strength and density may correlate and anti-correlate in turbulent interstellar gas within the fluctuations and this can introduce the dispersion of the mass-to-flux ratios within the turbulent volume. Each of the fluctuations provide a {\it transient} change of the pointwise magnetization. In the absence of other effects, e.g. related to the thermal instability, the de-correlation is reversible. In comparison, the ``reconnection diffusion'', similar to the ambipolar diffusion, deals with the {\it permanent} de-correlation of magnetic field and density. Acting alone, the ``reconnection diffusion'' increases entropy making magnetic field-density de-correlation irreversible.

What are the laws that govern \textit{magnetic field diffusion} in turbulent magnetized fluids? Could those affect our understanding of basic interstellar and star formation processes? These are the questions that we address in this paper.

In this study, we try to understand the diffusion of magnetic field in a couple of idealized models in the presence of turbulence. We explore setups both with and without gravity and compare the diffusion of magnetic field with that of a passive scalar. In the context of star formation an important issue that we will address is an
alternative way of decreasing the magnetic flux-to-mass ratio without appealing to ambipolar diffusion. We claim that since turbulence in astrophysics is really ubiquitous, our results should be widely applicable. We also perform simulations including turbulent heat diffusion, which allow us to compare results obtained with our code with those in the literature.

This paper is organized as follows. In Section 2, we draw the theoretical grounds about  fast magnetic reconnection. In Section 3, we describe the numerical code employed.  In Section 4, we present the results concerning the diffusion of magnetic field in a setup without external gravitational forces. In Section 5, we present the results of our experiments of diffusion of magnetic field in the presence of a gravitational field. In Section 6, we discuss our results and compare with previous works.
In Section 7, we discuss the accomplishments and limitations of our  present study. In Section 8, we discuss our findings in the context of strong turbulence theory, and finally in Section 9, we summarize our conclusions.
While our work is focused on the diffusion of magnetic fields, we address in the Appendix the heat transport 
in magnetized turbulent plasma.  We confirm with higher resolution the results in \citet{cho2003a} that the heat advection within turbulent flows in the presence of magnetic fields is very similar to that induced by hydrodynamic turbulence. 

\section{Fast magnetic reconnection in the presence of turbulence and magnetic diffusion due to reconnection}

The dynamical response of magnetic fields in turbulent fluids, as we discussed above, depends on the ability of magnetic fields to change their topology via reconnection. We know from observations that magnetic field reconnection may be both fast and slow. Indeed, a slow phase of reconnection is necessary in order to explain the accumulation of free energy associated with the magnetic flux that precedes eruptive flares in magnetized coronae. Thus it is important to identify the conditions for the reconnection to be fast. Different mechanisms prescribe different necessary requirements for this to happen.

The problem of magnetic reconnection is most frequently discussed in terms of solar flares. However, this is a general basic process underlying the dynamics of magnetized fluids in general. If the magnetic field lines in a turbulent fluid do not easily reconnect, the properties of the fluid should be dominated by intersecting magnetic flux tubes which are unable to pass through each other. Such fluids cannot be simulated with the existing codes as magnetic flux tubes readily reconnect in the numerical simulations which are currently very diffusive compared to the actual astrophysical flows.

\begin{figure}[!t]
 \begin{center}
\includegraphics[width=1.0 \columnwidth]{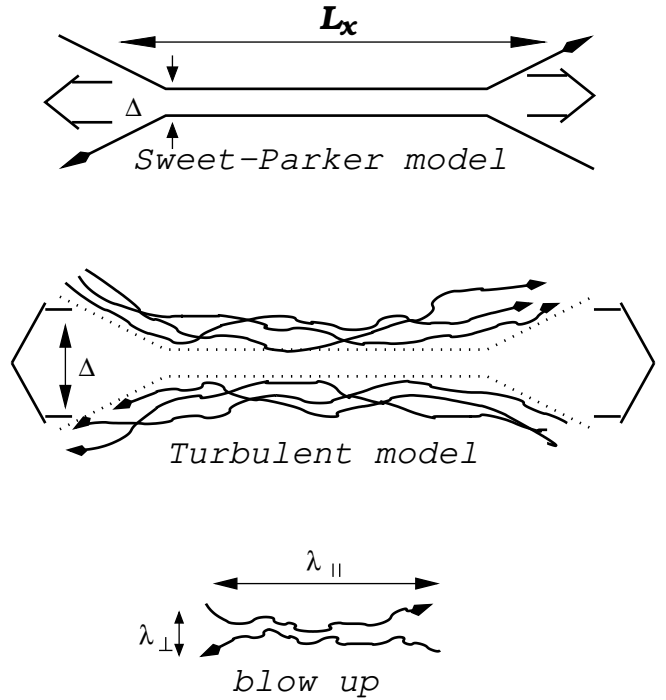}
\caption{{\it Upper plot}:
Sweet--Parker model of reconnection. The outflow
is limited by a thin slot $\Delta$, which is determined by Ohmic
diffusivity. The other scale is an astrophysical scale $L_{x} \gg \Delta$.
{\it Middle plot}: reconnection of weakly stochastic magnetic field according to
LV99. The model that accounts for the stochasticity
of magnetic field lines. The outflow is limited by the diffusion of
magnetic field lines, which depends on field line stochasticity.
{\it Low plot}: an individual small-scale reconnection region. The
reconnection over small patches of magnetic field determines the local
reconnection rate. The global reconnection rate is substantially larger
as many independent patches come together. From \citet{lazarian2004}.}
\label{recon1}
 \end{center}
\end{figure}

The famous Sweet-Parker model of reconnection (\citealt{sweet1958,parker1958}; see Figure~\ref{recon1}, upper panel) produces reconnection rates which are smaller than the Alfv\'en velocity by a square root of the
Lundquist number, i.e. by $S^{-1/2}\equiv (L_{x}V_A/\eta)^{-1/2}$, where $L_{x}$ in this case is the length of the current sheet. Astrophysical values of $S$ can be as large as $10^{15}$ or $10^{20}$, thus this scheme produces reconnection at a rate which is negligible for most of the astrophysical circumstances. If the Sweet--Parker were proven to be the only possible model of reconnection, it would have been possible to show that MHD numerical simulations do not have anything to do with real astrophysical fluids. Fortunately, faster schemes of reconnection are available.

The first model of fast reconnection proposed by \citet{petschek1964} assumed that magnetic fluxes get into contact not along the astrophysically large scales of $L_{x}$, but instead over a scale comparable to the resistive thickness $\delta$, forming a
distinct X-point, where magnetic field lines of the interacting fluxes converge at a sharp point to the reconnection spot. The stability of such a reconnection geometry in astrophysical situations is an open issue. At least for uniform resistivity, this configuration was proven to be unstable and to revert to a Sweet--Parker configuration \citep{biskamp1986,uzdensky2000}.

Recent years have been marked by the progress in understanding some of the key processes of reconnection in astrophysical plasmas. In particular, a substantial progress has been obtained by considering reconnection in the presence of the Hall-effect \citep{shay1998, shay2004}.
The condition for which the Hall-MHD term becomes important for the reconnection is that the ion skin depth $\delta_{\text{ion}}$ becomes comparable with the Sweet-Parker diffusion scale $\delta_{\text{SP}}$. The ion skin depth is a microscopic characteristic and it can be viewed as the gyroradius of an ion moving at the Alfv\'en speed, i.e. $\delta_{\text{ion}}=V_A/\omega_{ci}$, where $\omega_{ci}$ is the cyclotron frequency of an ion. For the parameters of the ISM (see Table~1 in \citealt{draine1998}), the reconnection is collisional (see further discussion in \citealt{yamada2006}).

To deal with both collisional and collisionless plasma \citet[henceforth LV99]{lazarian1999} proposed a model of fast reconnection in the presence of weak turbulence when magnetic field back-reaction is extremely important. We stress that the LV99 model does not assume that the magnetic field can be easily bent by fluid motions, which is a usual assumption to ``justify'' the {\it erroneous} concept of turbulent diffusivity in the mean field dynamo.

The middle and bottom panels of Figure~\ref{recon1} illustrate the key components of LV99 model\footnote{The cartoon in Figure~\ref{recon1} is an idealization of the reconnection process as the actual reconnection region also includes reconnected open loops of magnetic field moving oppositely to each other. Nevertheless, the cartoon properly reflects the role of the three-dimensionality of the reconnection process, the importance of small-scale reconnection events, and the increase of the outflow region compared to the Sweet--Parker scheme.}. The reconnection events happen on small scales $\lambda_{\|}$ where magnetic field lines get into contact. As the number of independent reconnection events that take place simultaneously is $L_{x}/\lambda_{\|}\gg 1$ the resulting reconnection speed is not limited by the speed of individual events on the scale $\lambda_{\|}$. Instead, the constraint on the reconnection speed comes from the thickness of the outflow reconnection region $\Delta$, which is determined by the magnetic field wandering in a turbulent fluid. The model is intrinsically three dimensional\footnote{Bidimensional numerical simulations of turbulent reconnection in \citet{kulpa-dybel2009} show that the reconnection is not fast in this case.} as both field wandering and simultaneous entry of many independent field patches, as shown in Figure~\ref{recon1}, are three-dimensional effects. In LV99 model the magnetic reconnection speed becomes comparable with $V_A$ when the scale of magnetic field wandering $\Delta$ becomes comparable with $L_{x}$.\footnote{Another process that is determined by magnetic field wandering is the diffusion of energetic particles perpendicular to the mean magnetic field. Indeed, the coefficient of diffusion perpendicular to the magnetic field in the Milky Way is just an order of unity less than the coefficient of diffusion parallel to the magnetic field (see \citealt{giacalone1999}, and referenes therein).}

For a quantitative description of the reconnection, one should adopt a model of MHD turbulence (see \citealt{iroshnikov1963, kraichnan1965, dobrowolny1980, shebalin1983, montgomery1981, higdon1984}). The most important for magnetic field wandering is the Alfv\'enic component. Adopting the \citet[henceforth GS95]{goldreich1995} scaling of the Alfv\'enic component of MHD turbulence extended to include the case of weak turbulence, LV99 predicted that the reconnection speed in a weakly turbulent magnetic field is
\begin{equation}
V_{R}=V_A (l/L_{x})^{1/2} (v_{l}/V_A)^2
\label{recon}
\end{equation}
where the level of turbulence is parameterized by the injection velocity; the combination $V_A (v_{l}/V_A)^2$
is the velocity of the largest strong turbulence eddies $V_{\text{strong}}$, i.e., the velocity at the scale at which the Alfv\'enic turbulence transfers from the weak to the strong regimes. Thus Equation~(\ref{recon}) can also be rewritten as $V_{R}=V_{\text{strong}} (l/L_{x})^{1/2}$; $v_{l}<V_A$ and $l$ is the turbulence injection scale.

The scaling predictions given by Equation~(\ref{recon}) have been tested successfully by three-dimensional MHD numerical simulations in \citet{kowal2009}. This stimulates us to adopt the LV99 model as a starting point for our discussion of magnetic reconnection.

How can $\lambda_{\|}$ be determined? In LV99 model, as many as $L_{x}^2/\lambda_{\bot}\lambda_{\|}$ localized reconnection events take place, each of which reconnects the flux at the rate $V_{\text{rec, local}}/\lambda_{\bot}$, where $V_{\text{rec, local}}$ is the velocity of local reconnection events at the scale $\lambda_{\|}$. The individual reconnection events contribute to the global reconnection rate, which in three dimensions becomes a factor of $L_{x}/\lambda_{\|}$ larger, i.e.,
\begin{equation}
V_{\text{rec,global}}\approx L_{x}/\lambda_{\|} V_{\text{rec, local}}.
\label{global}
\end{equation}

The local reconnection speed, conservatively assuming that the local events are happening at the Sweet--Parker rate, can be easily obtained by identifying the local resistive region $\delta_{\text{SP}}$ with $\lambda_{\bot}$ and using the relations between $\lambda_{\|}$ and $\lambda_{\bot}$ that follow from the MHD turbulence model. The corresponding calculations in LV99 provided the local reconnection rate $v_l S^{-1/4}$. Substituting this local reconnection rate in Equation~(\ref{global}) one estimates the global reconnection speed, {\it if this speed were limited by Ohmic resistivity}, which is larger than $V_A$ by a factor $S^{1/4}$. As a result, one has to conclude that the reconnection does not depend on resistivity.

The LV99 model is applicable to situations when plasma effects are included, e.g. the Hall-MHD effect, which increases effective resistivities for local reconnection events. While the latter point is difficult to test directly with existing plasma codes, e.g., with PIC codes, due to the necessity of simulating both plasma microphysics effects as well as macrophysical effects of magnetic turbulence, \citet{kowal2009} simulated the action of plasma effects by parameterizing them via anomalous resistivities. The values of such resistivities are a steep function of the separation between oppositely directed magnetic field lines, which also determines the current separating magnetic fluxes. With anomalous resistivities, the structure of the fractal current sheet of the turbulent reconnection changed substantially, but no significant changes of the reconnection rate were reported, which agrees well with the theoretical expectations of LV99. Within the model the explanation of this stems from the fact that the reconnection is already fast (i.e., independent of resistivity) even when small-scale reconnection events are mediated by the Ohmic resistivity, while the bottleneck for the reconnection process is provided by magnetic field wandering. Thus the increase of the local reconnection rate does not increase the global reconnection speed.

In this work, we address the problems which are relevant to the reconnection in a partially ionized, weakly turbulent gas. The corresponding model of reconnection was proposed in \citet[henceforth LVC04]{lazarian2004}. The extensive calculations summarized in Table 1 in LVC04 show that the reconnection for realistic circumstances varies from $0.1V_A$ to $0.03V_A$, i.e., is also fast, which should enable fast diffusion arising from turbulent motions.

The fact that magnetic fields reconnect fast in turbulent fluids ensures that the large-scale dynamics that we can reproduce well with numerical codes is not compromised by the difference in reconnection processes in the computer and in astrophysical flows. This motivates our present study in which we investigate diffusion processes in turbulent magnetized fluids via three-dimensional simulations. We feel that the issue of the dimension of the simulations is crucial for the turbulence dynamics  and the reconnection and, as a result, for our final results.

Visualization of turbulent diffusion of heat or any passive scalar field is easy within the GS95 model, which can be interpreted as a model of Kolmogorov cascade perpendicular to the local direction of the magnetic field (LV99 and more discussion in Section 8). The corresponding eddies are expected to advect heat similarly to the case of the hydrodynamic heat advection (see Appendix A). The corresponding visualization of the magnetic field diffusion is more involved. Every time that magnetic field lines intersect each other, they change their configuration draining free energy from the system. In the presence of self-gravity this may mean the escape of magnetic field, which is a ``light fluid'' from the self-gravitating gaseous ``heavy fluid''. Naturally, if the turbulence gets very strong the system gets unbound and then the mixing of magnetic field and gas, rather than their segregation is expected. In what follows we test these ideas numerically.

\section{MHD equations and the numerical code}

The systems studied numerically in this work is described by the resistive MHD equations, assuming an isothermal equation of state:

\begin{equation}
\frac{\partial \rho}{\partial t}  + \nabla \cdot \left( \rho \mathbf{u} \right)  = 0
\end{equation}
\begin{equation}
\rho \left( \frac{\partial}{\partial t} + \mathbf{u} \cdot \nabla \right) \mathbf{u} = - c_{s}^{2} \nabla \rho + (\nabla \times \mathbf{B}) \times \mathbf{B} - \rho \nabla \Psi + \mathbf{f}
\end{equation}
\begin{equation}
\frac{\partial \mathbf{B}}{\partial t} = \nabla \times \left( \mathbf{u} \times \mathbf{B} \right) + \eta _{\text{Ohm}} \nabla ^{2} \mathbf{B}
\end{equation}
plus the divergenceless condition for the magnetic field $\nabla \cdot \mathbf{B} = 0$. The spatial coordinates are given in units of a typical length $L_{*}$. The density $\rho$ is normalized by a reference density $\rho _{*}$, and the velocity field $\mathbf{u}$ by a reference velocity $U_{*}$. The constant sound speed $c_{s}$ is also given in units of $U_{*}$, and the magnetic field $\mathbf{B}$ is measured in units of $U_{*} \sqrt{4 \pi \rho _{*}}$. The uniform Ohmic resistivity $\eta _{\text{Ohm}}$ is given in units of $U_{*} L_{*}$. In our numerical calculations we will  use both non-zero and zero values of $\eta_{\text{Ohm}}$. In the latter case, the calculations will include only  numerical resistivity.  Time $t$ is measured in units of $L_{*} / U_{*}$. The external gravitational potential $\Psi$ is given in units of $U_{*}^{2}$. The source term $\mathbf{f}$ is a random force term responsible for injection of turbulence.

The above equations are solved inside a three-dimensional box with periodic boundary conditions. We use a
shock-capturing Godunov-type scheme with an HLL solver (see, for example, \citealt{kowal2007,falceta2008}). Time integration is performed with the Runge-Kutta method of second order. Unless we say explicitly the opposite, we assume $\eta _{\text{Ohm}} = 0$.

We employ an isotropic, non-helical, solenoidal, delta correlated in time forcing $\mathbf{f}$. This forcing acts in a thin shell around the wave number $k_{f} = 2.5 ( 2 \pi / L )$, that is, the scale of turbulence injection $l_{\text{inj}}$ that is about $2.5$ times smaller than the box size $L$ (in all the simulations, we choose $L=1$ in code units). In most of the experiments, the $rms$ velocity $V_{\text{rms}}$ induced by turbulence in the box is close to unity (in code units). Therefore, in these cases, the turnover time of the energy-carrying eddies (or the turbulent timescale $t_{\text{turb}}$) is $t_{\text{turb}} \sim l_{\text{inj}} / v_{\text{turb}} \sim (L/2.5) / V_{\text{rms}} \approx 0.4$ units of  time in code units.

We note that for our present purposes we use an one-fluid approximation, which does not include ambipolar diffusion. This choice is appropriate for approximating a fully ionized gas. One may argue that the code describes also the dynamics of partially ionized gas, but on the scales where ions and neutrals are strongly coupled, i.e,. on scales larger than the scale of ambipolar diffusion (see discussion in \citealt{lazarian2004}).

\section{Turbulent magnetic field diffusion in the absence of gravity}

 Observations of different regions of the diffuse ISM compiled by \citet{troland1986} indicate that magnetic fields and density are not straightforwardly correlated. These observations motivated \citet{heitsch2004} to perform 2.5-dimensional numerical calculations in the presence of both ambipolar diffusion and turbulence. The results in \citet{heitsch2004} also indicated de-correlation of magnetic field and density\footnote{We feel that the constrained geometry of the simulations  in \citet{heitsch2004} (where the magnetic field was assumed to be perpendicular to the plane of the two-dimensional turbulence, so that there were no reconnection) weakened the comparison of their set up with effects in the magnetized ISM, but this point is beyond the scope of our present discussion.} and one may wonder whether ambipolar diffusion is always required to de-correlate magnetic field and density or if, otherwise, to what extent the concept of ``turbulent ambipolar diffusion'' introduced in \citet{heitsch2004} is useful (see also \citealt{zweibel2002}). 
To address these issues we performed three-dimensional simulations of magnetic diffusion in the absence of ambipolar diffusion effects.

\subsection{Numerical Setup}

The magnetic field is assumed to have initially only the component in the $z$-direction. The initial configuration of the magnetic and density fields are:

\begin{equation}
B_{z} (x,y) = B_{0} + B_{1} \cos\left( \frac{2\pi}{L} x \right) \cos\left( \frac{2\pi}{L} y \right)
\end{equation}
\begin{eqnarray}
\rho (x,y) = & \rho _{0} - \frac{1}{c_{s}^{2}} \left\lbrace B_{0} B_{1} \cos\left( \frac{2\pi}{L} x \right) \cos\left( \frac{2\pi}{L} y \right) \right. \nonumber \\ & + 0.5 \left.\left[ B_{1} \cos\left( \frac{2\pi}{L} x \right)\cos \left( \frac{2\pi}{L} y \right) \right] ^{2} \right\rbrace \text{,}
\end{eqnarray}
where $(x,y)=(0,0)$ is the center of the $x,y$-plane. Boundary conditions are periodic.

\begin{figure*}[!hbt]
 \begin{center}
 \includegraphics[width=0.8 \textwidth]{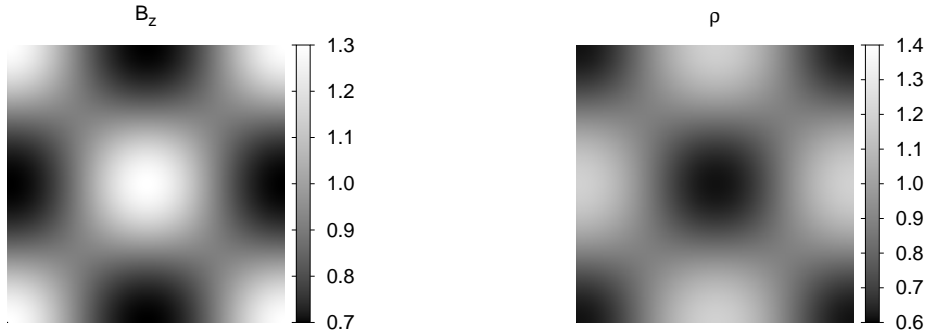}
 \caption{($x,y$)-plane showing the initial configuration of the $z$ component of the magnetic field $B_{z}$ ({\it left}) and the density distribution ({\it right}) for the model B2 (see Table \ref{tab:heitsch2}). The centers of the plots correspond to $(x,y)=(0,0)$.}
 \label{fig:heitsch_ilu1}
 \end{center}
\end{figure*}

This initial magnetic field configuration has an uniform component $B_{0}$ plus an harmonic perturbation of amplitude $B_{1}$. The density field is distributed in such a way that the gas pressure, given by the isothermal equation of state $p = c_{s}^{2} \rho$, equilibrates exactly the magnetic pressure, giving a magneto-hydrostatic solution. We choose the parameters $B_{1}=0.3$, $\rho _{0}=1$ and $c_{s}=1$ in all our simulations. Figure \ref{fig:heitsch_ilu1} illustrates these initial fields when $B_{0}=1.0$.

We should remark that the simulations presented in \citet{heitsch2004} do not start at the equilibrium, like ours. There is no pressure term in their equation for the evolution of the momentum of the ions to counterbalance the magnetic pressure. In addition, in their work the ion-density field is kept constant in time and space.

Another difference between our setup and that in \citet{heitsch2004} is that our parameter $B_{1}$ is assumed to be the same for all the models studied, and not a fraction of $B_{0}$. Also, the amplitude of the perturbation of the homogeneous component of the density field is a free parameter in \citet{heitsch2004}, while here it is constrained by the imposed equilibrium between gas and magnetic pressures.

In addition, we introduce a passive scalar field $\Phi$ initially identical to $B_{z}$. The parameters of our relevant simulations are presented in Table \ref{tab:heitsch2}.

\begin{table*}[!hbt]
\caption{Parameters of the Simulations in the Study of Turbulent Diffusion of Magnetic Flux without Gravity.}
\centering
\begin{tabular}{c c c c c c c c}
\hline \hline
Model	&	$B_{0}$	&	$l_{\text{inj}}$	&	$V_{\text{rms}}$ &	$t _{\text{turb}}$ &	$\eta _{\text{turb}}-B_{z}$ &	$\eta _{\text{turb}}-\Phi$ &	 Resolution \\
[0.5ex]
\hline
B1	&	$0.5$	&	$0.4$ &	$0.8$ &	$0.5$ &	$0.18(0.03)$ &	$0.19(0.02)$ &	$256^{3}$ \\
B2	&	$1.0$	&	$0.4$ &	$0.8$ &	$0.5$ &	$0.09(0.02)$ &	$0.14(0.02)$ &	$256^{3}$ \\
B3	&	$1.5$	&	$0.4$ &	$0.8$ &	$0.5$ &	$0.10(0.02)$ &	$0.08(0.02)$ &	$256^{3}$ \\
B4	&	$2.0$	&	$0.4$ &	$0.8$ &	$0.5$ &	$0.13(0.02)$ &	$0.11(0.02)$ &	$256^{3}$ \\
B2l	&	$1.0$	&	$0.4$ &	$0.8$ &	$0.5$ &	$0.15(0.02)$ &	$0.11(0.02)$ &	$128^{3}$ \\
B2h	&	$1.0$	&	$0.4$ &	$0.8$ &	$0.5$ &	$0.10(0.01)$ &	$0.13(0.02)$ &	$512^{3}$ \\
[1ex]
\hline
\end{tabular}
\label{tab:heitsch2}
\end{table*}

In these simulations we keep the random velocity approximately constant, i.e., $V_{\text{rms}} \approx 0.8$ for all the models after one time step. Therefore, all these models are slightly subsonic.

\subsection{Notation}

Hereafter, the quantities within brackets with  subscript ``$R=0.25L$'': $\left\langle \cdot \right\rangle _{R=0.25L}$, or simply ``$0.25$'': $\left\langle \cdot \right\rangle _{0.25}$ will denote averages inside a cylinder with main-axis in the $z$-direction centered in the computational box, with radius $R=0.25L$, while a subscript ``$z$'': $\left\langle \cdot \right\rangle _{z}$ will denote an average over the $z$-direction. An overbar means the average of some quantity inside the entire box.

\subsection{Results}

\begin{figure}[!hbt]
 \begin{center}
 \includegraphics[width=1.0 \columnwidth]{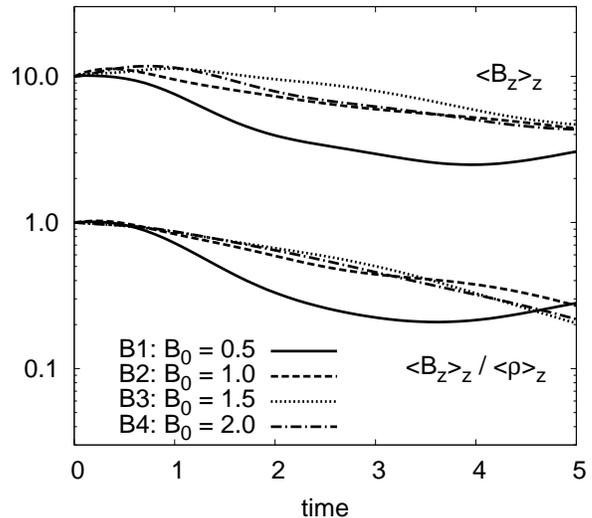}
 \caption{Evolution of the $rms$ amplitude of the Fourier modes $(k_{x},k_{y})=(\pm 1,\pm 1)$ of $\left\langle B_{z} \right\rangle _{z}$ (upper curves) and $\left\langle B_{z} \right\rangle _{z} / \left\langle \rho \right\rangle _{z}$ (lower curves). The curves for $\left\langle B_{z} \right\rangle _{z}$ were multiplied by a factor of $10$. All the curves were smoothed to make the visualization clearer.}
 \label{fig:heitsch8B}
 \end{center}
\end{figure}

Figure \ref{fig:heitsch8B} shows the evolution of the amplitude of the mode that is identical to the initial harmonic perturbation of the magnetic field (i.e., the $rms$ of the amplitude of the Fourier modes $(k_{x},k_{y})=(\pm 1,\pm 1)$), for $\left\langle B_{z} \right\rangle _{z}$ and
$\left\langle B_{z} \right\rangle _{z} / \left\langle \rho \right\rangle _{z}$. Most right plot in Figure \ref{fig:heitsch8} (Appendix) shows the evolution of the amplitude of the same mode for $\left\langle \Phi \right\rangle _{z}$ and $\left\langle \Phi \right\rangle _{z} / \left\langle \rho \right\rangle _{z}$. All the curves presented were smoothed in order to make the visualization clearer. We see that the decay of the magnetic field occurs at a  similar rate to that of the passive field. The mode decays nearly exponentially at roughly the same rate for most of the models. Only the model B1 ($B_{0}=0.5$) exhibits a higher decay rate. This may be due to the large scale field reversals that are common in super-Alfv\'enic turbulence. Table \ref{tab:heitsch2} shows the fitted values (and the uncertainty) of $\eta_{\text{turb}}$ in the curves corresponding to the evolution of the amplitude of the modes for $\left\langle B_{z} \right\rangle _{z}$. The fitted curve is $\exp \left\lbrace -k^{2} \eta_{\text{turb}} t \right\rbrace$, where $k^{2}=k_{x}^{2}+k_{y}^{2}=2$ is the square of the module of the corresponding wave-vector. We observe that the decay of the amplitude of the  modes of $\left\langle B_{z} \right\rangle _{z}$  is not continuous but saturates at a value that is naturally maintained by the turbulence (in Figure 3 it occurs after about $t=6$).

The diffusion of $B / \rho$ on large scales was also observed in \citet{heitsch2004} for  two-fluid simulations and there it was associated with the difference between the velocity field of the ions and neutrals, at small scales. However, here we observe a similar effect, but in one-fluid simulations, which is suggestive that turbulence rather than the details of the microphysics are responsible for the diffusion.

\begin{figure*}[!hbt]
 \begin{center}
 \includegraphics[width=1.0 \textwidth]{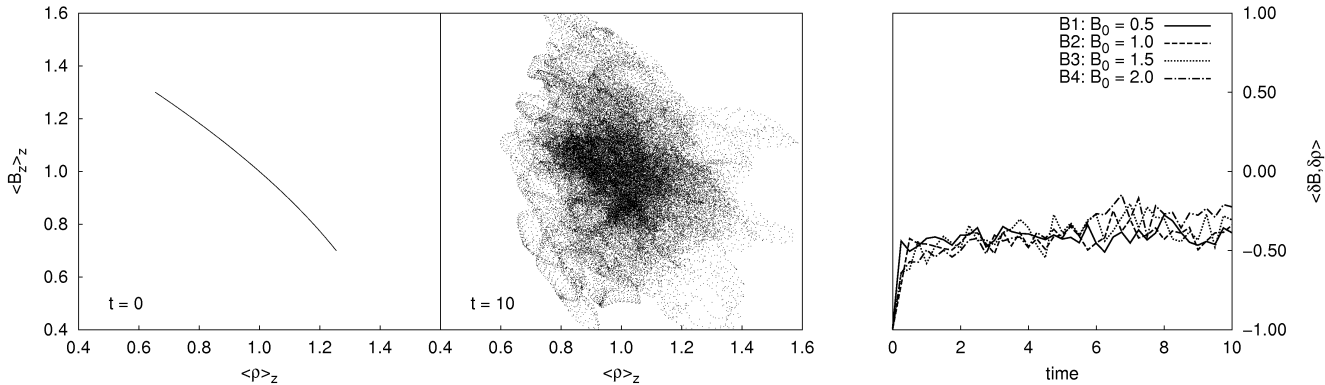}
 \caption{Distribution of $\left\langle \rho \right\rangle _{z}$ vs. $\left\langle B_{z} \right\rangle _{z}$ for model B2 (see Table \ref{tab:heitsch2}), at $t=0$ (left) and $t=10$ (center). \textit{Right}: correlation between fluctuations of the strength of the magnetic field ($\delta B$) and density ($\delta \rho$). }
 \label{fig:heitsch11}
 \end{center}
\end{figure*}

Left and center panels of Figure \ref{fig:heitsch11} shows the distribution of $\left\langle \rho \right\rangle _{z}$ versus $\left\langle B_{z} \right\rangle _{z}$ for the model B2 ($B_{0}=1.0$) at the initial configuration ($t=0$) and after $10$ time steps. We see in this projected view that the initial magnetic flux-to-mass relation is quickly spread and, in contrast with the $\Phi--\rho$ distribution (see Figure \ref{fig:heitsch11b}, in the Appendix), we do not see any tendency for the magnetic field and density to become correlated.

To give a quantitative measure of the evolution of the flux-to-mass relation in the models, let us consider $\left\langle \delta B, \delta \rho \right\rangle$, the correlation between fluctuations of the magnetic field $\delta B$ and density $\delta \rho$, defined by
\begin{equation}
\left\langle \delta B, \delta \rho \right\rangle =
\frac{\int  (B - \bar{B}) (\rho - \bar{\rho}) d^{3}x}{ \sqrt{\int (B - \bar{B})^{2} d^{3}x} \sqrt{\int (\rho - \bar{\rho})^{2} d^{3}x} } \text{.}
\label{eqn:correlation}
\end{equation}
Right panel of Figure \ref{fig:heitsch11} shows the evolution of $\left\langle \delta B, \delta \rho \right\rangle$ (see right side of Figure \ref{fig:heitsch11b} for the evolution of $\left\langle \delta \Phi, \delta \rho \right\rangle$ which is similarly defined). Differently from the passive scalar field that quickly becomes correlated to the density field, the magnetic field keeps a residual anti-correlation with it.

A more careful analysis of our results indicates that the correlation between magnetic field intensity and density depends on the Mach number $M_{s}$. For example, when
we calculate the correlation $\left\langle \delta B, \delta \rho \right\rangle$ using the turbulent models of Table \ref{tab:diffusion1} (study of diffusion of passive scalar fields, see Appendix A), we find weak positive correlations for the supersonic models and negative correlations for the subsonic ones. These correlations increase with $M_{s}$. Thus, the anti-correlation detected in Figure \ref{fig:heitsch11} can be due to the slightly subsonic regime of the turbulence. These correlations and anti-correlations at this level cannot be excluded by the observational data as discussed, e.g., in \citet{troland1986}. We shall address this issue in more detail elsewhere.

To summarize, the results of Figures \ref{fig:heitsch8B} and \ref{fig:heitsch11} suggest that the turbulence can substantially change the flux-to-mass ratio $B/\rho$ without any effect of ambipolar diffusion. The diffusion of the magnetic flux occurs in a rate similar to the rate of the turbulent diffusion of heat (passive scalar), even for sub-Alfv\'enic turbulence.

As remarked in Section 2, we should  emphasize that the efficient turbulent diffusion of magnetic field that we are observing in the simulations above is due to fast magnetic reconnection because otherwise, if the tangled magnetic lines by turbulence were not reconnecting, then they would be behaving like a Jello-type substance and this would make  the diffusive transport of magnetic flux very inefficient (contrary to what is observed in the simulations). The issue of magnetic reconnection was avoided in \citet{heitsch2004} due to the settings in which magnetic field was assumed perpendicular to the plane of the fluid motions. Magnetic reconnection, however, is an effect that is present within any realistic three-dimensional setup.

\subsection{Effects of Resolution on the Results}

\begin{figure}[!hbt]
 \begin{center}
 \includegraphics[width=1.0 \columnwidth]{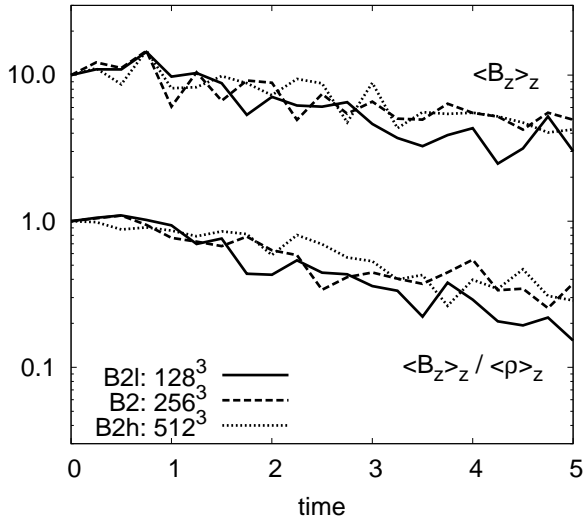}
 \caption{Comparison between models of different resolution: B2, B2l, and B2h (Table \ref{tab:heitsch2}). It presents the same quantities as in Figure \ref{fig:heitsch8B}.}
 \label{fig:heitsch_resol}
 \end{center}
\end{figure}

To convince the reader that the above results are not being affected by numerical effects, we ran one of the models (model B2) with increased and decreased resolutions (models B2h and B2l, respectively, see Table \ref{tab:heitsch2}). Figure \ref{fig:heitsch_resol} compares the same quantities presented in Figure \ref{fig:heitsch8B} for these models. We do not observe significant difference between them. Thus, we can expect that the results presented for the models with resolution of $256^3$ are robust.

\section{``Reconnection diffusion'' in the presence of gravity}

The ``reconnection diffusion'' of magnetic field in the absence of gravity can represent the magnetic field dynamics in diffuse interstellar gas where cloud self-gravity is not important. In molecular clouds, clumps and accretion disks, the diffusion of magnetic field should happen in the presence of gravity. We study this process below.

\subsection{Numerical Approach}

In order to get an insight into the magnetic field diffusion in a turbulent fluid immersed in a gravitational potential, we have performed experiments in the presence of a gravitational potential with cylindric symmetry  $\Psi$, given in cylindrical coordinates $(R, \phi, z)$ by:
\begin{equation}
\Psi (R \leq R_{\text{max}}) =  - \frac{A}{R + R_{*}}
\label{eq:pot1}
\end{equation}
\begin{equation}
\Psi (R > R_{\text{max}}) =  - \frac{A}{R_{\text{max}} + R_{*}}
\label{eq:pot2}
\end{equation}
where $R=0$ is the center of the $(x,y)$-plane, and we fixed $R_{*} = 0.1 L$ and $R_{\text{max}} = 0.45 L$ ($L=1$ is the size of the computational box, as remarked in Section 3). We assume a relatively high value of $R_{*}$ in order to limit the values of the gravitational force and prevent the system to be initially Parker--Rayleigh--Taylor unstable. We assume an outer cut-off $R_{\text{max}}$ on the gravitational force to ensure the cylindrical symmetry while using periodic boundary conditions.

In one class of experiments, we start the simulation with a magneto-hydrostatic equilibrium with $\beta = P_{\text{gas}} / P_{\text{mag}} = c_{s}^{2} \rho / (B^2/8 \pi)$ constant. The initial density and magnetic fields are, respectively,
\begin{equation}
\rho (R) = \rho _0 \exp \left\lbrace (\Psi(R_{\text{max}}) - \Psi(R)) / c_{s}^{2} (1 + \beta^{-1}) \right\rbrace
\end{equation}
\begin{equation}
B_{z} (R) = c_{s} \sqrt{2 \beta^{-1} \rho (R)} \text{.}
\end{equation}

\begin{figure*}[!hbt]
 \begin{center}
 \includegraphics[width=1.0 \textwidth]{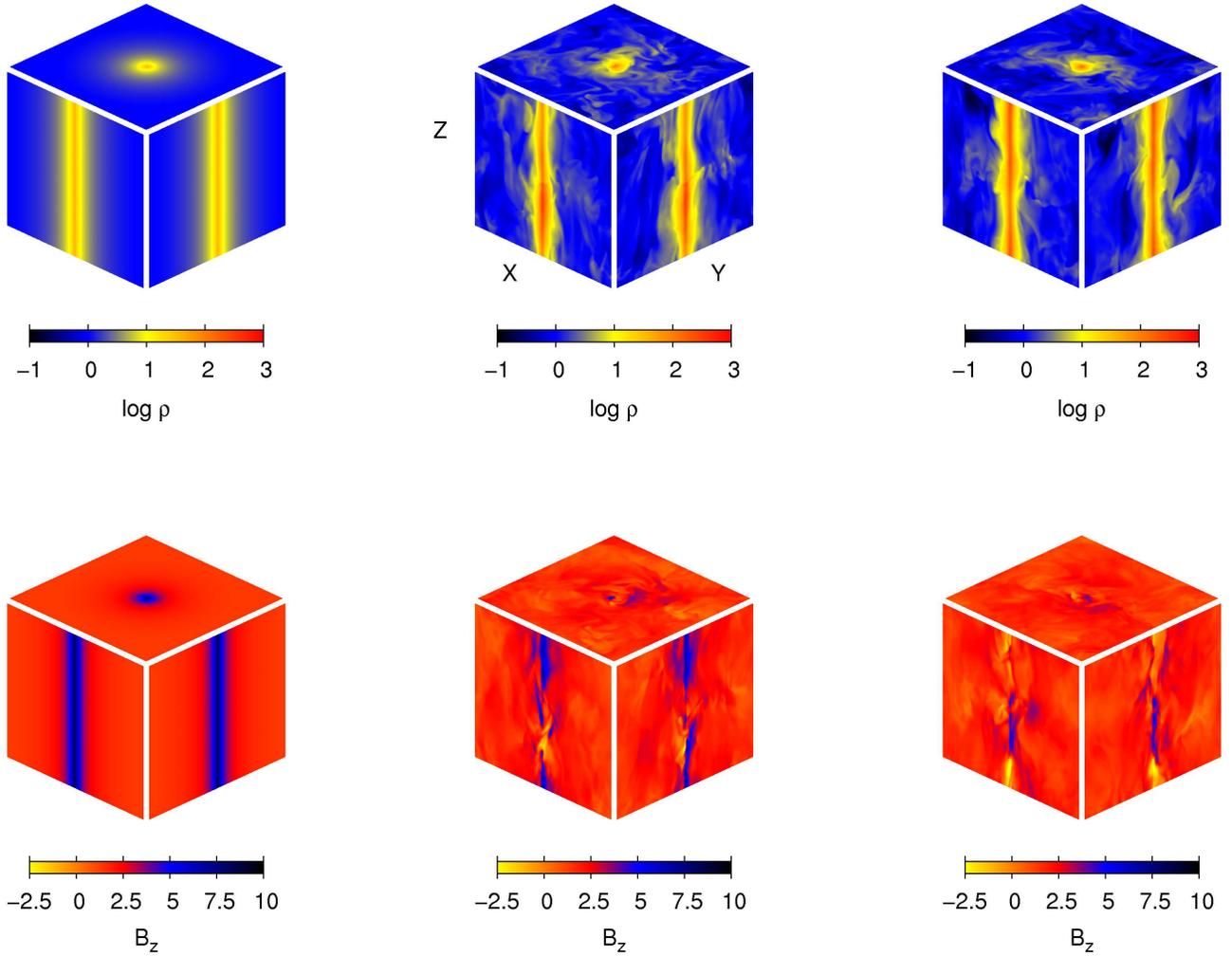}
 \caption{Model C2 (see Table \ref{tab:cylind}). \textit{Top row}: logarithm of the density field; \textit{bottom row}: $B_{z}$ component of the magnetic field. \textit{Left column}: central $xy$, $xz$, and $yz$ slices of the system projected on the respective walls of the cubic computational domain, in $t=0$; \textit{middle and right columns}: the same for $t=3$ (middle) and $t=8$ (right). }
 \label{fig:grav_ilu}
 \end{center}
\end{figure*}

Figure \ref{fig:grav_ilu} illustrates this initial configuration for one of the studied models (model C2, see Table \ref{tab:cylind}).

We restricted our experiments to the trans-sonic case $c_{s}=1$ (in most of the experiments, we keep $V_{\text{rms}} \approx 0.8$, see Table \ref{tab:cylind}). We also fixed $\rho _{0} = 1$. The turbulence is injected at $t=0$ and we follow the evolution of $\langle B_{z} \rangle _{R}$ and $\langle \rho \rangle _{R}$ for eight time steps. Table \ref{tab:cylind} lists the parameters used for these experiments.  $\bar{\rho}$ and $\bar{B_{z}}$ represent the average of the density and magnetic field over the entire box. $V_{A,i}$ refers to the initial Alfv\'en speed of the system. The $rms$ velocity of the system $V_{\text{rms}}$ is measured after the turbulence is well-developed.

\begin{table*}[!hbt]
\begin{center}
\caption{Parameters for the Models with Gravity Starting at Magneto-hydrostatic Equilibrium with Initial Constant $\beta$. }
\centering
\begin{tabular}{c c c c c c c c c c c}
\hline \hline
Model	&	$\beta$\tablenotemark{1}	&	$V_{A,i}$	&	$A$\tablenotemark{2}	&	$\bar{\rho}$ &	$\bar{B_{z}}$ &	 $V_{\text{rms}}$	&	$t_{\text{turb}}$ &	 $\eta_{\text{turb}}$ &	$\eta_{\text{turb}} / V_{\text{rms}} l_{\text{inj}}$ &	Resolution \\
[0.5ex]
\hline
C1	&	$1.0$	&	$1.4$	&	$0.6$	&	$1.26$ &	$1.59$ &	$0.8$	&	$0.5$ &	$\lessapprox 0.005$ &	 $\lessapprox 0.015$ &	 $256^{3}$ \\
C2	&	$1.0$	&	$1.4$	&	$0.9$	&	$1.52$ &	$1.74$ &	$0.8$	&	$0.5$ &	$\approx 0.01$ &	$\approx 0.03$ &	 $256^{3}$ \\
C3	&	$1.0$	&	$1.4$	&	$1.2$	&	$1.95$ &	$1.98$ &	$0.8$	&	$0.5$ &	$\approx 0.03$ &	$\approx 0.09$ &	 $256^{3}$ \\
C4	&	$1.0$	&	$1.4$	&	$0.9$	&	$1.52$ &	$1.74$ &	$1.4$	&	$0.3$ &	$\approx 0.10-0.20$ &	 $\approx 0.18-0.36$ &	 $256^{3}$ \\
C5	&	$1.0$	&	$1.4$	&	$0.9$	&	$1.52$ &	$1.74$ &	$2.0$	&	$0.2$ &	$\gtrapprox 0.30$ &	$\gtrapprox 0.37$ &	 $256^{3}$ \\
C6	&	$3.3$	&	$0.8$	&	$0.9$	&	$2.40$ &	$1.20$ &	$0.8$	&	$0.5$ &	$\approx 0.02$ &	$\approx 0.06$ &	 $256^{3}$ \\
C7	&	$0.3$	&	$2.4$	&	$0.9$	&	$1.18$ &	$2.66$ &	$0.8$	&	$0.5$ &	$\approx 0.01$ &	$\approx 0.03$ &	 $256^{3}$ \\
C2l	&	$1.0$	&	$1.4$	&	$0.9$	&	$1.52$ &	$1.74$ &	$0.8$	&	$0.5$ &	... &	... &	 $128^{3}$ \\
C2h	&	$1.0$	&	$1.4$	&	$0.9$	&	$1.52$ &	$1.74$ &	$0.8$	&	$0.5$ &	... &	... &	 $512^{3}$ \\
[1ex]
\hline
\end{tabular}
\label{tab:cylind}

\vspace{1mm}
\begin{flushleft}
$^{1}${Initial $\beta$ parameter for the plasma: $\beta = P_{\text{gas}} / P_{\text{mag}}$ }

$^{2}${The parameter $A$ for the strength of gravity (see Equations 9 and 10) is given in units of $c_s^2 L$.}
\end{flushleft}

\end{center}
\end{table*}

We have also performed experiments starting out of equilibrium, with homogeneous fields: the system starts in free fall. We leave the system to evolve for eight time steps applying turbulence from the very beginning. For a comparison, we have also performed these experiments without turbulence. The initial uniform magnetic field is parallel to the $z$-direction for these models. Table \ref{tab:cylind2} lists the parameters for these runs. The listed values of $\beta$ refer to the initial conditions.

\begin{table}[!hbt]
\caption{Parameters for the Models with Gravity Starting Out-of-equilibrium, with Initially Homogeneous Fields.}
\centering
\begin{tabular}{c c c c c c c c c c}
\hline \hline
Model	&	$\beta$	&	$V_{A,i}$	&	$A$	&	$\bar{\rho}$ &	$\bar{B_{z}}$ &	$V_{\text{turb}}$	&	Resolution \\
[0.5ex]
\hline
D1	&	$1.0$	&	$1.4$	&	$0.9$	&	$1.0$ &	$1.41$ &	$0.8$	&	$256^{3}$ \\
D1a	&	$1.0$	&	$1.4$	&	$0.9$	&	$1.0$ &	$1.41$ &	$0.0$	&	$256^{3}$ \\
D2	&	$3.3$	&	$0.8$	&	$0.9$	&	$1.0$ &	$0.77$ &	$0.8$	&	$256^{3}$ \\
D2a	&	$3.3$	&	$0.8$	&	$0.9$	&	$1.0$ &	$0.77$ &	$0.0$	&	$256^{3}$ \\
D3	&	$0.3$	&	$2.4$	&	$0.9$	&	$1.0$ &	$2.45$ &	$0.8$	&	$256^{3}$ \\
D3a	&	$0.3$	&	$2.4$	&	$0.9$	&	$1.0$ &	$2.45$ &	$0.0$	&	$256^{3}$ \\
D1l	&	$1.0$	&	$1.4$	&	$0.9$	&	$1.0$ &	$1.41$ &	$0.8$	&	$128^{3}$ \\
D1h	&	$1.0$	&	$1.4$	&	$0.9$	&	$1.0$ &	$1.41$ &	$0.8$	&	$512^{3}$ \\
[1ex]
\hline
\end{tabular}
\label{tab:cylind2}
\end{table}

Concerning the diffusion of the magnetic field, in order to provide a quantitative comparison between the models, we have also performed simulations with similar initial conditions to the models presented in Table \ref{tab:cylind}, but without turbulence and with the explicit presence of Ohmic diffusivity $\eta_{\text{Ohm}}$ in the induction equation. As these models have perfect symmetry in the $z$-direction, we simulated only a plane cutting the $z$-axis, that is, they are 2.5-dimensional simulations. We use a resolution comparable to the turbulent three-dimensional models. Table \ref{tab:cylind3} lists the parameters for theses runs. We simulated a three-dimensional model equivalent to the model E7r1, and we found exact agreement in the time evolution of the magnetic flux distribution (not shown). Therefore, we can believe that in this case these 2.5-dimensional simulations give results which are equivalent to three-dimensional simulations.

\begin{table}[!hbt]
\caption{Parameters for the 2.5-dimensional Resistive Models with Gravity Starting with Magneto-hydrostatic Equilibrium and Constant $\beta$.}
\centering
\begin{tabular}{c c c c c}
\hline \hline
Model	&	$\beta$	&	$A$	&	$\eta_{\text{Ohm}}$	&	Resolution \\
[0.5ex]
\hline
E1r0	&	$1.0$	&	$0.6$	&	$0.005$	&	$256^{2}$ \\
E1r1	&	$1.0$	&	$0.6$	&	$0.01$	&	$256^{2}$ \\
E2r1	&	$1.0$	&	$0.9$	&	$0.01$	&	$256^{2}$ \\
E2r2	&	$1.0$	&	$0.9$	&	$0.02$	&	$256^{2}$ \\
E2r3	&	$1.0$	&	$0.9$	&	$0.03$	&	$256^{2}$ \\
E2r4	&	$1.0$	&	$0.9$	&	$0.05$	&	$256^{2}$ \\
E3r1	&	$1.0$	&	$1.2$	&	$0.01$	&	$256^{2}$ \\
E3r2	&	$1.0$	&	$1.2$	&	$0.02$	&	$256^{2}$ \\
E3r3	&	$1.0$	&	$1.2$	&	$0.03$	&	$256^{2}$ \\
E4r2	&	$1.0$	&	$0.9$	&	$0.10$	&	$256^{2}$ \\
E4r3	&	$1.0$	&	$0.9$	&	$0.20$	&	$256^{2}$ \\
E5r3	&	$1.0$	&	$0.9$	&	$0.30$	&	$256^{2}$ \\
E6r1	&	$3.3$	&	$0.9$	&	$0.01$	&	$256^{2}$ \\
E6r2	&	$3.3$	&	$0.9$	&	$0.02$	&	$256^{2}$ \\
E6r3	&	$3.3$	&	$0.9$	&	$0.03$	&	$256^{2}$ \\
E7r1	&	$0.3$	&	$0.9$	&	$0.01$	&	$256^{2}$ \\
[1ex]
\hline
\end{tabular}
\label{tab:cylind3}
\end{table}

\subsection{Results}

\subsubsection{Evolution of the Equilibrium Distribution}

\begin{figure*}[!hbt]
 \begin{center}
 \includegraphics[width=1.0 \textwidth]{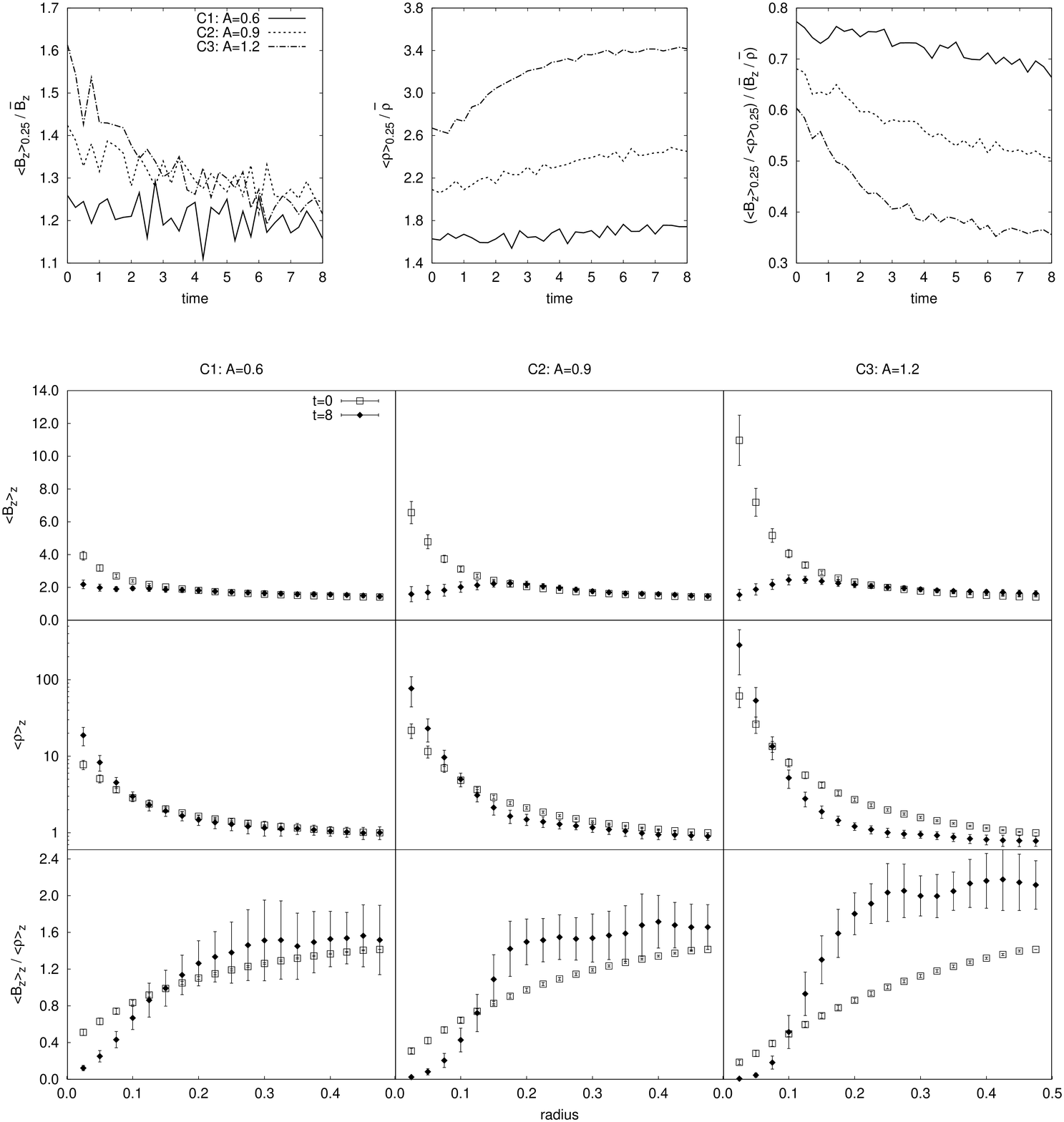}
 \caption{Evolution of the equilibrium models for different gravitational potential. The top row shows the time evolution of $\left\langle B_{z} \right\rangle _{0.25} / \bar{B_{z}}$ (\textit{left}), $\left\langle \rho \right\rangle _{0.25} / \bar{\rho}$ (\textit{middle}), and $(\left\langle B_{z} \right\rangle _{0.25} / \left\langle \rho \right\rangle _{0.25}) / (\bar{\rho} / \bar{B_{z}})$ (\textit{right}). The other plots show the radial profile of $\left\langle B_{z} \right\rangle _{z}$ (\textit{upper panels}), $\left\langle \rho \right\rangle _{z}$ (\textit{middle panels}), and $\left\langle B_{z} \right\rangle _{z} / \left\langle \rho \right\rangle _{z}$ (\textit{bottom panels}) for the different values of $A$ in $t=0$ (magneto-hydrostatic solution with $\beta$ constant, see Table \ref{tab:cylind}) and $t=8$. Error bars show the standard deviation. All models have initial $\beta=1.0$. }
 \label{fig:grav2}
 \end{center}
\end{figure*}

Top row of Figure \ref{fig:grav2} shows the evolution of $\left\langle B_{z} \right\rangle _{0.25}$ (\textit{left}), $\left\langle \rho \right\rangle _{0.25}$ (\textit{middle}), and $\left\langle B_{z} \right\rangle _{0.25} / \left\langle \rho \right\rangle _{0.25}$ (\textit{right}), normalized by the respective characteristic values inside the box ($\bar{B_{z}}$, $\bar{\rho}$ and $\bar{B_{z}}/\bar{\rho}$), for the models C1, C2, and C3 ($\beta=1$).
We compare the evolution of these quantities for different strengths of gravity $A$, maintaining the other parameters
identical. The central magnetic flux reduces faster the higher the value of $A$. The flux-to-mass ratio has similar
behavior. The other plots of Figure \ref{fig:grav2} show the profile of the quantities
$\left\langle B_{z} \right\rangle _{z}$ (\textit{upper panels}),
$\left\langle \rho \right\rangle_{z} $ (\textit{middle panels}),
and $\left\langle B_{z} \right\rangle _{z} / \left\langle \rho \right\rangle _{z}$ (\textit{bottom panels})
along the radius $R$, each column corresponding to a different value of $A$ both for $t=0$
(in magneto-hydrostatic equilibrium and constant $\beta$) and for $t=8$.
We see  the deepest decay of the magnetic flux toward the central region for the highest value of $A$ at $t=8$.

\begin{figure*}[!hbt]
 \begin{center}
 \includegraphics[width=1.0 \textwidth]{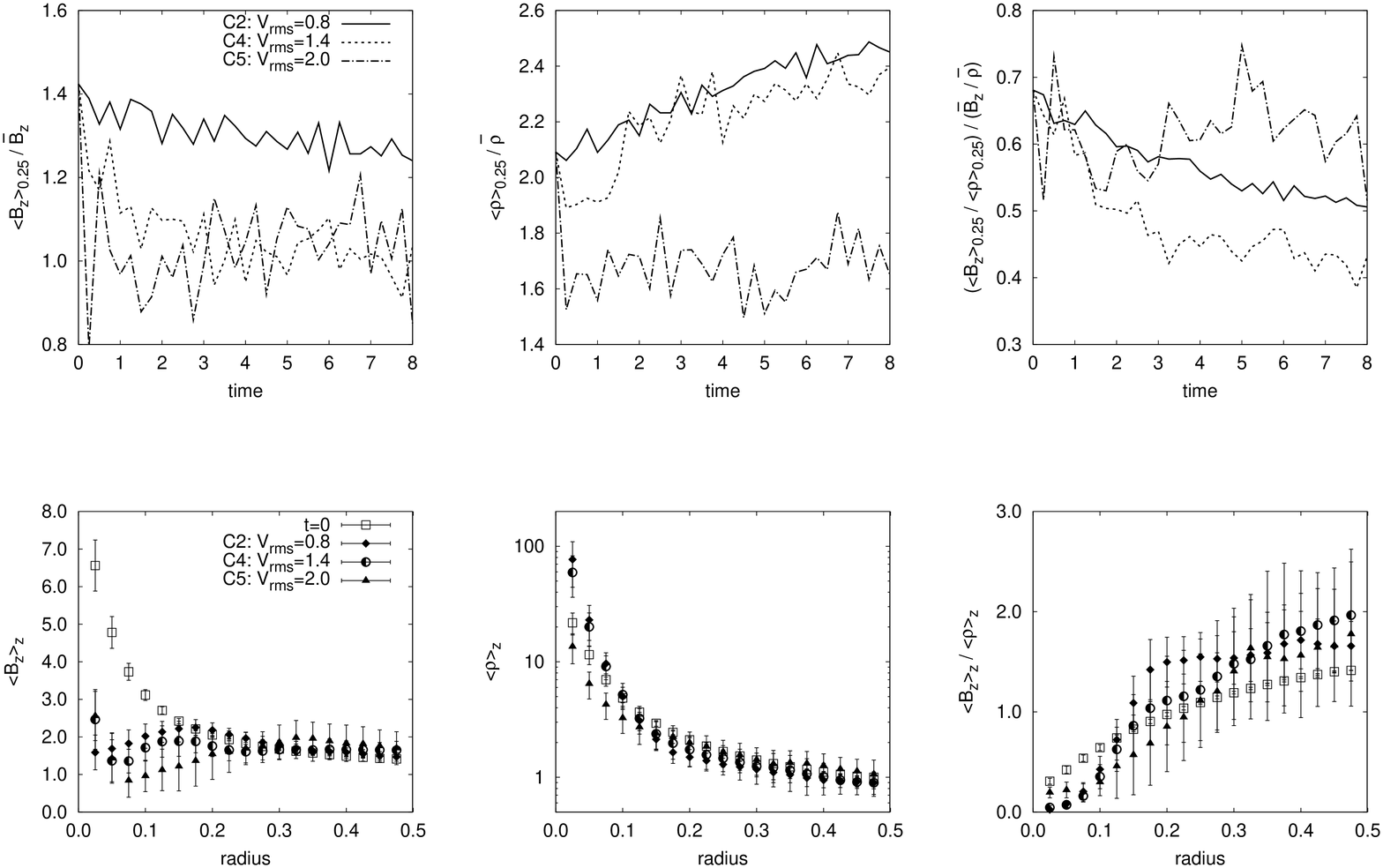}
 \caption{Evolution of the equilibrium models for different turbulent driving. The top row shows the time evolution of $\left\langle B_{z} \right\rangle _{0.25} / \bar{B_{z}}$ (\textit{left}), $\left\langle \rho \right\rangle _{0.25} / \bar{\rho}$ (\textit{middle}), and $(\left\langle B_{z} \right\rangle _{0.25} / \left\langle \rho \right\rangle _{0.25}) / (\bar{\rho} / \bar{B_{z}})$ (\textit{right}). The bottom row shows the radial profile of $\left\langle B_{z} \right\rangle _{z}$ (\textit{left}), $\left\langle \rho \right\rangle _{z}$ (\textit{middle}), and $\left\langle B_{z} \right\rangle _{z} / \left\langle \rho \right\rangle _{z}$ (\textit{right}) for each value of the turbulent velocity $V_{\text{rms}}$, in $t=0$ (magneto-hydrostatic solution with $\beta$ constant) and $t=8$. Error bars show the standard deviation. All models have initial $\beta=1.0$. See Table \ref{tab:cylind}.}
 \label{fig:grav1}
 \end{center}
\end{figure*}

In Figure \ref{fig:grav1}, we compare the rate of the ``reconnection diffusion''
when we change the turbulent velocity and maintain the other parameters identical as in  models C2, C4 and C5.
An inspection of the left panel shows that the central magnetic flux $\left\langle B_{z} \right\rangle _{0.25}$
decreases faster for the two highest turbulent velocities.
Fluctuations are higher in the cases with higher velocity. The central density however,
gets smaller for the highest turbulent velocity.
This is explained by the fact that the dynamic pressure is higher for the largest velocities.
The central flux-to-mass ratio $\left\langle B_{z} \right\rangle _{0.25} / \left\langle \rho \right\rangle _{0.25}$
decays for the two smallest velocities. However, for the largest velocity, it is not clear if this ratio
decreases or not. Looking at the middle graph of Figure \ref{fig:grav1} (\textit{bottom row}), we see that the central density
decreases for the highest forcing. This is indicative that the turbulence driving
overcomes the gravitational potential making the system less bound.

\begin{figure*}[!hbt]
 \begin{center}
 \includegraphics[width=1.0 \textwidth]{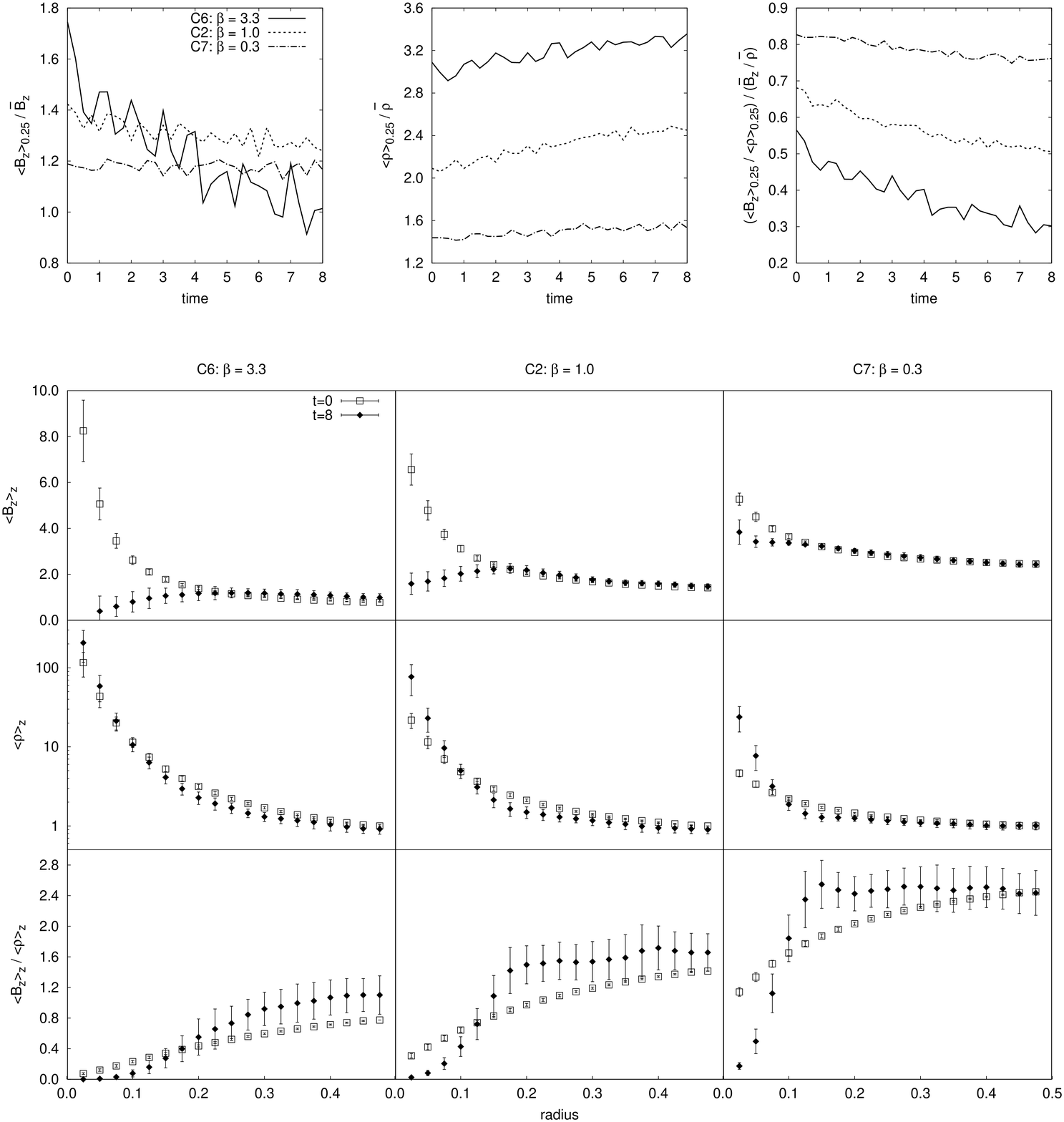}
 \caption{Evolution of the equilibrium models for different degrees of magnetization (plasma $\beta=P_{\text{gas}}/P_{\text{mag}}$). The top row shows the time evolution of $\left\langle B_{z} \right\rangle _{0.25} / \bar{B_{z}}$ (\textit{left}), $\left\langle \rho \right\rangle _{0.25} / \bar{\rho}$ (\textit{middle}), and $(\left\langle B_{z} \right\rangle _{0.25} / \left\langle \rho \right\rangle _{0.25}) / (\bar{\rho} / \bar{B_{z}})$ (\textit{right}). The other plots show the radial profile of $\left\langle B_{z} \right\rangle _{z}$ (\textit{upper panels}), $\left\langle \rho \right\rangle _{z}$ (\textit{middle panels}), and $\left\langle B_{z} \right\rangle _{z} / \left\langle \rho \right\rangle _{z}$ (\textit{bottom panels}) for each value of $\beta$, in $t=0$ (magneto-hydrostatic solution with $\beta$ constant) and $t=8$. Error bars show the standard deviation of the data. See Table \ref{tab:cylind}.}
 \label{fig:grav3}
 \end{center}
\end{figure*}

Top row of Figure \ref{fig:grav3} compares the evolution of $\left\langle B_{z} \right\rangle _{0.25}$ (\textit{left}), $\left\langle \rho \right\rangle _{0.25}$ (\textit{middle}), and $\left\langle B_{z} \right\rangle _{0.25} / \left\langle \rho \right\rangle _{0.25}$ (\textit{right}),
normalized by the  characteristic average values inside 
the box ($\bar{B_{z}}$, $\bar{\rho}$ and $\bar{B_{z}}/\bar{\rho}$), for models C2, C6, and C7 with different $\beta$. 
Both the central magnetic flux and the flux-to-mass ratio decreases faster 
for the less magnetized model ($\beta=3.3$). The other plots of Figure \ref{fig:grav3} show the 
radial profile of the 
quantities 
$\left\langle B_{z} \right\rangle _{z}$ (\textit{upper panels}), $\left\langle \rho \right\rangle _{z}$ (\textit{middle panels}), and $\left\langle B_{z} \right\rangle _{z} / \left\langle \rho \right\rangle _{z}$ (\textit{bottom panels})
for each model. We can again observe a lower value of the flux in the central region
 (relative to the external regions) for the highest values of $\beta$ at the time step $t=8$. 
 The contrast between the central and the more external values for the flux-to-mass ratio is 
 quite different for the three models, being higher for the more magnetized models. 
 This is expected, as turbulence brings the system in the state of minimal energy. 
 The effect of varying magnetization in some sense is analogous to the effect of varying gravity. 
 The equilibrium flux-to-mass ratio is larger in both the case of 
 higher gravity and higher magnetization. The physics is simple, 
 the lighter fluid (magnetic field) gets segregated from the heavier fluid (gas).

All in all, we clearly see that turbulence substantially influences the quasi-static evolution of magnetized gas in the gravitational potential. The system in the presence of turbulence relaxes faster to its minimum potential energy state. This explains the change of the flux-to-mass ratio, which for years was a problem to deal with invoking ambipolar diffusion.

\subsubsection{Equilibrium Models: Comparison of Magnetic Diffusivity and Resistivity Effects}

In terms of the removal of the magnetic field from quasi-static clouds,
does the effect of ``reconnection diffusion'' act similar to the effect of diffusion induced by resistivity?  To address this question, we have performed a series of simulations with enhanced Ohmic resistivity (see models of Table \ref{tab:cylind3}).

\begin{figure*}[!hbt]
 \begin{center}
 \includegraphics[width=1.0 \textwidth]{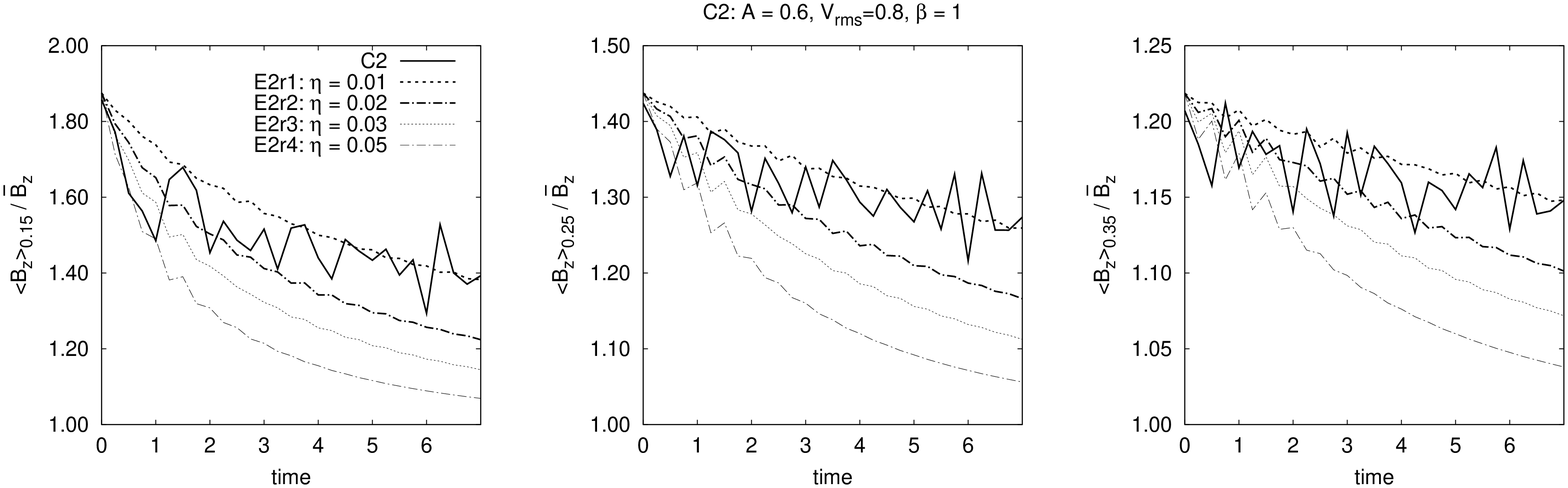}
 \caption{Comparison between the model C2 (turbulent diffusivity) and resistive models without turbulence (see Table \ref{tab:cylind3}). All the cases have analogous parameters. }
 \label{fig:ohmic1}
 \end{center}
\end{figure*}

In Figure \ref{fig:ohmic1}, we compare the evolution of $\left\langle B_{z} \right\rangle _{R}$ (at different radius) for model C2 of Table \ref{tab:cylind} with similar resistive models without turbulence of Table \ref{tab:cylind3}, with different values of Ohmic diffusivity $\eta_{\text{Ohm}}$. The decay seems initially faster and comparable with the highest value of $\eta_{\text{Ohm}}$ ($\eta_{\text{Ohm}}=0.05$). But after this initial phase, the turbulent model (C2) seems to have a behavior similar to the resistive models with $\eta_{\text{Ohm}}$ between $0.01$ and $0.02$.

Figure \ref{fig:ohmic2} (Appendix) compares the turbulent models C1, C3, C4, C5, C6, and C7 of Table \ref{tab:cylind} with similar resistive models of Table \ref{tab:cylind3}.
After roughly one time step,
the model C1 (weaker gravitational field) seems to be consistent with a
value of $\eta_{\text{Ohm}}$ between $0.005$ or lesser, while the model C3
(stronger gravitational field) seems to be consistent with $\eta_{\text{Ohm}} \approx 0.03$.
These results show that the effective turbulent magnetic diffusivity is sensitive
to the strength of the gravitational field.

 The resistive simulations with increasing $\eta_{\text{Ohm}}$ values are more comparable with models with increasing turbulent velocity.
 The model C4 (with smaller turbulent velocity) seems to be consistent with the resistive model with $\eta_{\text{Ohm}}=0.10$. The model C5 (with larger turbulent velocity) seems to be  more comparable with the model 
 with $\eta_{\text{Ohm}} \sim 0.30$ (or higher), however, in this case we cannot associate a representative value of $\eta_{\text{Ohm}}$ due to the very quick diffusion which occurs even before $t=1$, when the turbulence becomes well developed.

The turbulent curve for the less magnetized model (C6)
seems to follow the resistive curve with $\eta_{\text{Ohm}}=0.02$, while the more magnetized model (C7)
is comparable to the resistive model with $\eta_{\text{Ohm}} \approx 0.01$.
This result indicates that the effective turbulent diffusivity is also sensitive
to the strength of the magnetic field.

In summary, the results above indicate a correspondence between the two different effects. In other words, the turbulent magnetic diffusion may mimic the effects of Ohmic  diffusion of magnetic fields in gravitating clouds. However, we should keep in mind that the physics of turbulent diffusion and Ohmic resistivity is different. Thus this analogy should not be overstated.

\subsubsection{Evolution of Non-equilibrium Models}

\begin{figure*}[!hbt]
 \begin{center}
 \includegraphics[width=1.0 \textwidth]{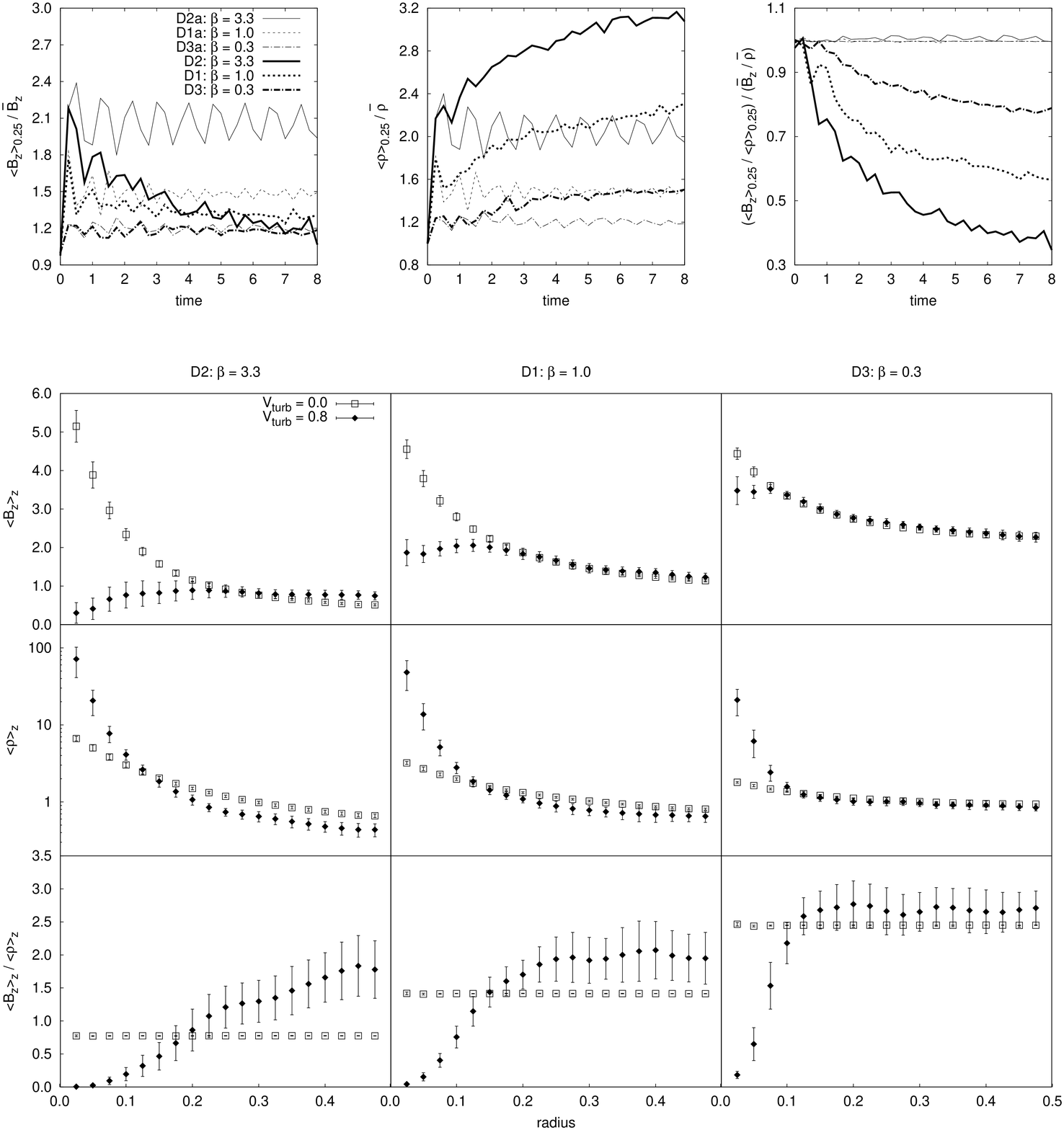}
 \caption{Evolution of models which start in non-equilibrium. The top row shows the time evolution of $\left\langle B_{z} \right\rangle _{0.25} / \bar{B_{z}}$ (\textit{left}), $\left\langle \rho \right\rangle _{0.25} / \bar{\rho}$ (\textit{middle}), and $(\left\langle B_{z} \right\rangle _{0.25} / \left\langle \rho \right\rangle _{0.25}) / (\bar{\rho} / \bar{B_{z}})$ (\textit{right}), for runs with (thick lines) and without (thin lines) injection of turbulence. The other plots show the radial profile of $\left\langle B_{z} \right\rangle _{z}$ (\textit{upper panels}), $\left\langle \rho \right\rangle _{z}$ (\textit{middle}), and $\left\langle B_{z} \right\rangle _{z} / \left\langle \rho \right\rangle _{z}$ (\textit{right}) for different values of $\beta$, at $t=8$, for runs with and without turbulence. Error bars show the standard deviation. See Table \ref{tab:cylind2}. }
 \label{fig:grav_freefall}
 \end{center}
\end{figure*}

Figure \ref{fig:grav_freefall} shows the same set of comparisons as in Figure \ref{fig:grav3} for the  models D1, D2, and D3
of Table \ref{tab:cylind2} --- these models have started out of the equilibrium with a homogeneous density and magnetic field
in a free fall system. Besides the runs with turbulence, we also present, for comparison,
the evolution for the systems without turbulence (models D1a, D2a, and D3a). The strong oscillations seen in the evolution of the central
magnetic flux and density for these models (which are more pronounced in the models without turbulence)
are acoustic oscillations, since the time for the virialization of these systems is larger than the simulated period.
We note that the initial flux-to-mass ratio does not change in the cases without turbulence.
We also observe similar trends as in Figure \ref{fig:grav3}: the higher the value of $\beta$,
the faster the decrease of the central magnetic flux relative to the mean flux into the box.
We also note that the radial profile of the flux-to-mass ratio for the turbulent models
 crosses the mean value for the models without turbulence
at nearly the same radius. This is due to the fact that the effective gravity potential in all these simulations acts up to
this radius approximately.

This set of simulations shows that the change of mass-to-flux ratio can happen at the time scale of the gravitational collapse of the system and therefore, turbulent diffusion of magnetic field is applicable also to dynamic situations, e.g., to the formation of supercritical cores.

\subsection{Effects of Resolution on the Results}

As in the case of the study presented in Section 4, we would like to know how the results shown in this section are sensitive to changes in resolution. Again, we ran some models employing higher resolution and we inspected the changes in the results concerned.

\begin{figure*}[!hbt]
 \begin{center}
 \includegraphics[width=1.0 \textwidth]{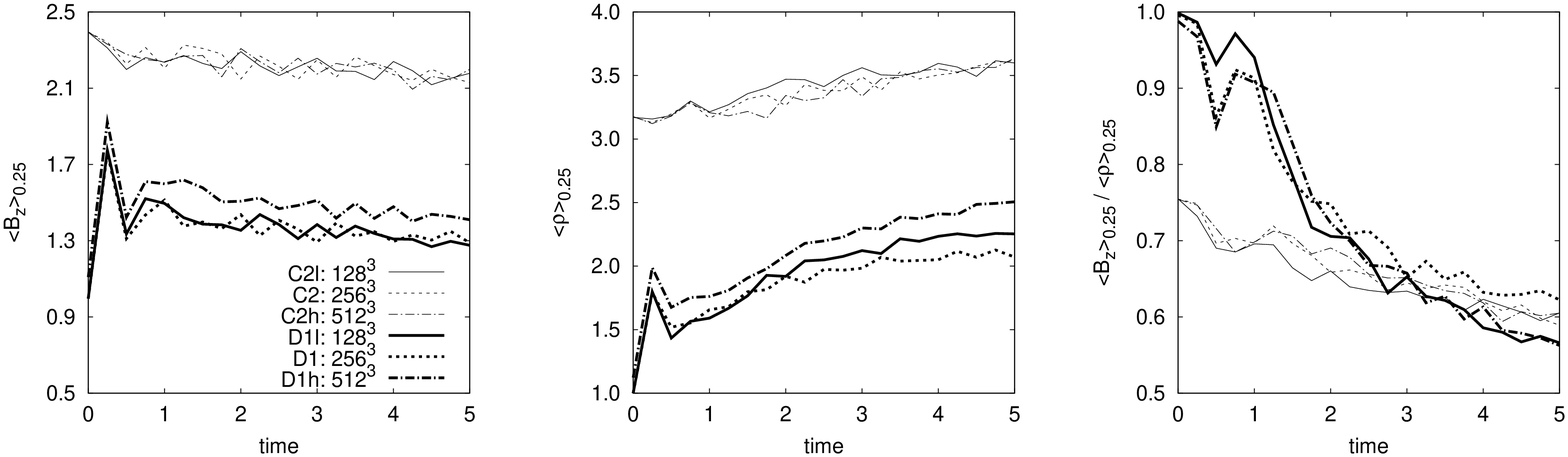}
 \caption{Comparison of the time evolution of $\left\langle B_{z} \right\rangle _{0.25}$ (\textit{left}), $\left\langle \rho \right\rangle _{0.25}$ (\textit{middle}), and $\left\langle B_{z} \right\rangle _{0.25} / \left\langle \rho \right\rangle _{0.25}$ (\textit{right}) between models with different resolutions: C2, C2l, C2h (Table \ref{tab:cylind}) and D2, D2l, D2h (Table \ref{tab:cylind2}).}
 \label{fig:grav_resol}
 \end{center}
\end{figure*}

Figure \ref{fig:grav_resol} compares the evolution of some of the quantities studied through this section for models C2l and C2h --- which are identical to C2, except by the lower (C2l) and higher resolution (C2h, see Table \ref{tab:cylind}). It shows no significant difference between these models. Figure \ref{fig:grav_resol} also depicts the evolution of the same quantities for the models D1, D1l, and D1h. Both models have the same parameters as in model D1, but model D1h (D1l) has higher (lower) resolution (see Table \ref{tab:cylind2}). Again, we see no disagreement between the models.

Therefore,  the results presented in this section are not expected to change with an increase in resolution.

\subsection{Magnetic Field Expulsion Revealed}

Both in the case of equilibrium and non-equilibrium we observe a substantial change of the mass-to-flux ratio. Even our experiments with no turbulence injection confirm that this process arises from the action of turbulence. As a result, in all the cases with gravity the turbulence allows magnetic field to escape from the dense core which is being formed in the center of the gravitational potential.

\section{Discussion of the results: relations to earlier studies}

Through this work we have performed the comparison of our results with the study by \citet{heitsch2004}. Below, we provide yet another outlook of the connection of that study with the present paper. We also discuss the work by \citet{shu2006}, which was the initial motivation of our study of the diffusion of magnetic field in the presence of gravity.

\subsection{Comparison with \citet{heitsch2004}: Ambipolar Diffusion Versus Turbulence and 2.5-dimensional Versus Three-dimensional}

In view of the astrophysical implications, the comparison between our results and those of \citet{heitsch2004} calls for the discussion on how ambipolar diffusion and turbulence interact to affect the magnetic field diffusivity. In particular, \citet{heitsch2004} claim that a new process ``turbulent ambipolar diffusion'' (see also \citealt{zweibel2002}) acts to induce fast magnetic diffusivity.

At the same time, our results do not seem to exhibit less magnetic diffusivity than those of \citet{heitsch2004} in spite of the fact that we do not have ambipolar diffusion. How can this be understood? We propose the following explanation. In the absence of ambipolar diffusion, the turbulence propagates to smaller scales making small-scale interactions possible. On the other hand, ambipolar diffusion affects the turbulence, increasing the damping scale. As a result, the ambipolar diffusion acts in two ways, in one to increase the small-scale diffusivity of the magnetic field, in another is to decrease the turbulent small-scale diffusivity and these effects essentially compensate each other\footnote{A possible point of confusion is related to the difference of the physical scales involved. If one associates the scale of the reconnection with the thickness of the Sweet--Parker layer, then, indeed,
the ambipolar diffusion scale is much larger and therefore the reconnection scale gets irrelevant. However, within the LV99 model of reconnection, the scale of reconnection is associated with the scale of magnetic field wandering. The corresponding scale depends on the turbulent velocity and is not small.}.

In other words, if we approximate the turbulent diffusivity by $(1/3) V_{\text{inj}} L_{\text{inj}}$, where $V_{\text{inj}}$ and $L_{\text{inj}}$ are the injection velocity and the injection scale for strong MHD turbulence (see LV99, \citealt{lazarian2006}), respectively, the ambipolar diffusivity acting on small scales will not play any role and the diffusivity will be purely ``turbulent''. If, however, the ambipolar diffusion coefficient is larger than $V_{\text{inj}} L_{\text{inj}}$, then the Reynolds number of the steered flow may become small for strong MHD turbulence to exist and the diffusion is purely ambipolar in this case. We might speculate that this leaves little, if any, parameter space for the ``turbulent ambipolar diffusion'' when turbulence and ambipolar diffusion synergetically enhance diffusivity, acting in unison. This point should be tested by three-dimensional two-fluid simulations exhibiting both ambipolar diffusion and turbulence.

In view of our findings one may ask whether it is surprising to observe the ``reconnection diffusion'' of magnetic field being independent of ambipolar diffusion. We can appeal to the fact well known in hydrodynamics, namely, that in a turbulent fluid the diffusion of a passive contaminant does not depend on the microscopic diffusivity. In the case of high microscopic diffusivity, the turbulence provides mixing down to a scale $l_1$ at which the microscopic diffusivity both, suppresses the cascade and ensures efficient diffusivity of the contaminant. In the case of low microscopic diffusivity, turbulent mixing happens down to a scale $l_2\ll l_1$, which ensures that even low microscopic diffusivity is sufficient to provide efficient diffusion. In both cases the total effective diffusivity of the contaminant is turbulent, i.e. is given by the product of the turbulent injection scale and the turbulent velocity. This analogy is not directly applicable to ambipolar diffusion, as this  is a special type of diffusion and magnetic fields are different from passive contaminants. However, we believe that our results show that to some extent the concept of turbulent diffusion developed in hydrodynamics carries over (due to fast reconnection) to magnetized fluid. 

\subsection{Transient De-correlation of Density and Magnetic Field}

Magneto-sonic waves are known to create transient changes of the density and magnetic field correlation. In the case of turbulence the situation is less clear, but the research in the field suggests that the decomposition of the turbulent motions into basic MHD modes is meaningful and justified even for high amplitude motions (\citealt{cho2003a}). Thus the claim in \citet{passot2003} that even in the limit of ideal MHD, turbulence can {\it transiently} affect the magnetic field and density correlations is justified. However, the process discussed in this paper is different in the sense that the de-correlation we describe here is {\it permanent} and it will not disappear if the turbulence dissipates. In a sense, as we showed above, ``reconnection diffusion'' is similar to the ambipolar and Ohmic diffusion. It is a dissipative diffusion process, which does require non-zero resistivity, although this resistivity can be infinitesimally small for the LV99 model of fast reconnection in the presence of turbulence (see Section 2).

\subsection{Relation to \citet{shu2006}: Fast Removal of Magnetic Flux During Star Formation}

As discussed in \citet{shu2006}, the sufficiency of the ambipolar diffusion efficiency for explaining observational data of accreting proto-stars is questionable. At the same time, they found that the required dissipation is about 4 orders of magnitude larger than the expected Ohmic dissipation. Thus they appealed to the hyper-resistivity concept in order to explain the higher dissipation of magnetic field.

We feel, however, that the hyper-resistivity idea is poorly justified (see criticism of it in \citealt{lazarian2004} and \citealt{kowal2009}). At the same time, fast three-dimensional ``reconnection diffusion'' can provide the magnetic diffusivity that is adequate for fast removing of the magnetic flux. This is what, in fact, was demonstrated in the present set of numerical simulations.

It is worth mentioning that, unlike the actual Ohmic diffusivity, ``reconnection diffusion'' does not transfer the magnetic energy directly into heat. The lion share of the energy is being released in the form of kinetic energy, driving turbulence (see LV99). If the system is initially laminar, this potentially result in flares of reconnection and the corresponding diffusivity. This is in agreement with LV99 scheme where a more intensive turbulence should induce more intensive turbulent energy injection and lead to the unstable feeding of the energy of the deformed magnetic field. However, the discussion of this effect is beyond the scope of the present paper.

Similar to \citet{shu2006}, we expect to observe the heating of the media. Indeed, although we do not expect to have Ohmic heating, the kinetic energy released due to magnetic reconnection is dissipated locally and therefore we expect to observe heating in the medium. Our setup for gravity can be seen as a toy model representing the situation in \citet{shu2006}. In the broad sense, our work confirms that a process of magnetic field diffusion that does not rely on ambipolar diffusion is efficient.

We accept that our setup assuming an axial gravitational field is a very simple and ignores complications that could arise from using a nearly spherical potential of the self-gravitating cloud. The periodic boundary conditions give super-stability to the system, and do not allow inflow (or outflow) of material/magnetic field as we expect in a more realistic accretion process. However, our experiments can give us qualitative insights. They show that the turbulent diffusion of the magnetic field can remove magnetic flux from the central region, leading to a lower flux-to-mass ratio in regions of higher gravity compared with that of lower gravity.

We chose parameters to the simulations such that the system is not initially unstable to the Parker--Rayleigh--Taylor (PRT) instability. Although the PRT instability could be present in real accretion systems and could help to remove magnetic field from the core of gravitational systems, its presence would make the interpretation of the results more difficult and we wanted to analyze only the turbulence role in the removal of magnetic flux. However, it is possible that this instability had been also acting due to local changes of parameters due to the turbulent motion. To ensure that the transport of magnetic flux is being caused by injection of turbulence only, we stopped the injection after a few time steps in some experiments and left the system to evolve. When we did this, the changes in the profile of the magnetic field and the other quantities stopped.

We showed that the higher the strength of the gravitational force, the lower the flux-to-mass ratio is  in the central region (compared with the mean value in the computational domain). This could be understood in terms of the potential energy of the system. When the gravitational potential well is deeper, more energetically  favorable is the pile up of matter near the center of gravity, reducing the total potential energy of the system. When the turbulence is increased, there is an initial trend to remove more magnetic flux from the center (and consequently more inflow of matter into the center), but for the highest value of the turbulent velocity in our experiments, there is a trend to remove material (together with magnetic flux) from the center, reducing the role of the gravity, due to the fact that the gravitational energy became small compared to the kinetic energy of the system. Our results also showed that when the gas is less magnetized (higher $\beta$, or higher values of the Alfv\'enic Mach number $M_{A}$), the reconnection diffusion of magnetic flux is more effective, but the central flux-to-mass ratio relative to external regions is  smaller for more magnetized models (low $\beta$), compared to less magnetized models. That is, the contrast $B / \rho$ between the inner and outer radius is higher for lower $\beta$ (or $M_{A}$).

If the turbulent diffusivity of magnetic field may explain the results in \citet{shu2006}, one may wonder whether one can remove magnetic field by this way not only from the class of systems studied by \citet{shu2006}, but also from less dense systems.
For instance, it is frequently assumed that only ambipolar diffusion is important for the evolution of subcritical magnetized clouds \citep{tassis2005}. Our study indicates that this conclusion may be altered in the presence of turbulence.  This point, however, requires further  careful study, which is beyond the scope of the present paper.
In the future, we intend to study a more realistic model, e.g., with open boundary conditions and more realistic gravitational potentials.

\section{Turbulent magnetic diffusion and turbulence theory}

The concept of ``reconnection diffusion'' is related to the LV99 model of fast reconnection  which makes use of the model of strong turbulence proposed by \citet[henceforth GS95]{goldreich1995} the turbulence is being injected at the large scales with the injection velocity $V_{\text{inj}}$ equal to the Alfv\'en velocity $V_A$ (see \citealt{cho2003b} for a review). The turbulent eddies mix up magnetic field mostly in the direction perpendicular to the local magnetic field thus forming a Kolmogorov-type picture in terms of perpendicular motions. Naturally, these eddies are as efficient as hydrodynamic eddies are expected in terms of heat advection. One also can visualize how such eddies can induce magnetic field diffusion.

It is important to note that the GS95 model deals with motions with respect to the {\it local} rather than mean magnetic field. Indeed, it is natural that the motions of the parcel of fluid are affected only by the magnetic field of the parcel and of the near vicinity, i.e., by local fields. At the same time,  in the reference frame of the mean field, the local magnetic fields of different parcels vary substantially. Thus we do not expect to see a substantial anisotropy of the heat advection when $V_{\text{inj}}\sim V_A$.

It was noted in LV99 that one can talk about turbulent eddies perpendicular to the magnetic field only if the magnetic field can reconnect fast. The rates of reconnection predicted in LV99 ensured that the magnetic field changes topology over one eddy turnover period. If the reconnection were slow, the magnetic fields would form progressively complex structures consisting of unresolved knots, which would invalidate the GS95 model. The response of such a fluid to mechanical perturbations would be similar to ``Jello'', making the turbulence-sponsored diffusion of magnetic field and heat impossible.

What happens  when $V_{\text{inj}}<V_A$?  In this case the turbulence at large scales is weak and therefore magnetic field mixing is reduced. Thus one may expect a partial suppression of magnetic diffusivity. However, as turbulence cascades the strength of interactions increases and at a scale $L_{\text{inj}} (V_{\text{inj}}/V_A)^2$ the turbulence gets strong. According to \citet{lazarian2006}, the diffusivity in this regime decreases by the ratio of $(V_{\text{inj}}/V_A)^3$, with the eddies of strong turbulence playing a critical role in the process. When we compare the turbulent diffusivity $\eta_{\text{turb}}$ estimated for the sub-Alfv\'enic models described in Table \ref{tab:cylind} (see Section 5) with $L_{\text{inj}} V_{\text{turb}} (V_{\text{turb}}/V_{A})^3$, we find that the values are roughly consistent with the predictions of \citet{lazarian2006}, although a more detailed study is required in this regard. For instance, we know that \citet{lazarian2006} theory was not intended for high Mach number turbulence.

All in all, we believe that the high diffusivity that we observe is related to the properties of strong magnetic turbulence. While the latter is still a theory which is  subject to intensive study (see \citealt{boldyrev2005,boldyrev2006,beresnyak2006,beresnyak2009a,beresnyak2009b,gogoberidze2007}), we believe that for the purpose of describing magnetic and heat diffusion the existing theory and the present model catch all the essential phenomena.

\section{Accomplishments and limitations of the present study}

\subsection{Major Findings}

This paper presents several sets of simulations which deal with magnetic diffusion in turbulent fluids. 
Comparing our result on magnetic diffusion and that of heat, we see many similarities in these two processes.
Our numerical testing in the paper would not make sense if the astrophysical reconnection were slow. Indeed, the major criticism that can be directed to the work of turbulent diffusion of heat by \citet{cho2003a} is that reconnection in their numerical simulations was fast due to high numerical diffusivity. With the confirmation of the LV99 model of turbulent reconnection by \citet{kowal2009} one may claim that astrophysical reconnection is also generally fast and the differences between the computer simulations and astrophysical flows are not so dramatic as far as the reconnection is concerned.

The most important part of our study is the removal of magnetic fields from gravitationally bounded systems (see Section 5). Generally speaking, this is what one can expect on the energetic grounds. Magnetic field can be identified with a light fluid which is not affected by gravity, while the matter tends to fall into the gravitational potential\footnote{As a matter of fact, in our low $\beta$ simulations, which we did not include in the paper, we see clear signatures of the PRT instability.}. Turbulence in the presence of magnetic reconnection helps ``shaking off'' matter from magnetic fields. In our simulations the gravitational energy was larger than the turbulent energy. In the case when the opposite is true, the system is expected to get unbounded with turbulence mixing magnetic field in the same way it does in the absence of gravity (see Section 4).

It is important to note that in Section 5 we obtained the segregation of magnetic field and matter both in the case when we started with equilibrium distribution and in the case when the system was performing a free fall. In the case of non-equilibrium initial conditions the amount of flux removed from the forming dense core is substantially larger than in the case of the equilibrium magnetic field/density configurations (compare Figures \ref{fig:grav3} and \ref{fig:grav_freefall}). Nevertheless, the flux removal happens fast, essentially in one turnover of the turbulent eddies. In comparison, the effect of numerical diffusion for the flux removal in our simulations is marginal, and this is testified by the constant flux-to-mass ratio obtained in the simulations without turbulence (see Figure \ref{fig:grav_freefall}).

What is the physical picture corresponding to our findings? In the absence of gravity turbulence mixes up\footnote{This mixing for Alfv\'enic modes happens mostly perpendicular to the local magnetic field for sub-Alfv\'enic and trans-Alfv\'enic turbulence (LV99). For super-Alfv\'enic turbulence the mixing is essentially hydrodynamic at large scales and the picture with motions perpendicular to the local magnetic field direction is restored at small scales (see discussion in \citealt{lazarian2004}).} flux tubes with different magnetic flux-to-mass ratios decreasing the difference in this ratio. In the presence of gravity, however, it is energetically advantageous of flux tubes at the center of the gravitational potential to increase the mass-to-flux ratio. This process is enabled in highly conducting fluid by turbulence which induces ``reconnection diffusion''. 

\subsection{Applicability of the Results}

The diffusion of magnetic field in our numerical runs exhibits a few interesting features. First of all, according to Figure \ref{fig:heitsch11} one may expect to see a broad distribution of magnetic field intensity with density. This seems to be consistent with the measurements of magnetic field strength in diffuse media \citep{troland1986}.

The situation gets even more intriguing as we discuss magnetic field diffusion in the gravitational potential. It is tempting to apply these results to star formation process (see studies by \citealt{leao2009}). There, molecular clouds are known to be either magnetically supercritical or magnetically subcritical (see \citealt{mestel1985}). If, however, the magnetic flux can be removed from the gravitating turbulent cloud in a timescale of about an eddy turnover time, then the difference between clouds with different initial magnetization becomes less important. The initially subcritical turbulent clouds can lose their magnetic flux via the turbulent diffusion to become supercritical.

An important point of the turbulent diffusion of the magnetic field is that it does not require gas to be weakly ionized, which is the requirement of the action of the ambipolar diffusion. Therefore, one may expect to observe gravitational collapse even of the highly ionized gas.

\subsection{Magnetic Field Reconnection and Different Stages of Star Formation}

``Reconnection diffusion'' seems to be a fundamental process that accompanies all the stages of star formation.
Our work shows (Section 4) that three-dimensional diffusion of magnetic field provides a wide distribution of the mass-to-flux ratios with some of the fluctuations having this ratio rather high. We believe that the diffusion of magnetic field described here is one the reasons for creation of zones of super-Alf\'enic turbulence even for sub-Alfv\'enic driving (see \citealt{burkhart2009}).

The regions of density concentration get gravitationally bound. One can associate such regions with GMCs. These entities are known to be highly turbulent and turbulent diffusion will proceed within them, providing a hierarchy of self-gravitating zones with different density and different mass-to-flux ratios. Some of those zones may be subcritical in terms of magnetic field and some of them may be supercritical. In subcritical magnetic cores the turbulent diffusion may proceed quasi-statically as we described in Section 5.2.1. and in the supercritical cores the turbulent diffusion may proceed as we described in Section 5.2.3. In both cases, we expect the removal of magnetic field from the self-gravitating cores. This process proceeds all the time, including the stage of the accretion disks.

Indeed, in the turbulent scenario of star formation it usually is assumed that at the initial stages of star formation the concentration of material happens due to gas moving along magnetic field lines \citep{mestel1984}. This one-dimensional process requires rather long times of the accumulation of material. In contrast, the ``reconnection diffusion'' allows for the much faster three-dimensional accumulation.

What is the relative role of the ambipolar diffusion and the ``reconnection diffusion''. This issue requires further studies. It is clear from the study by \citet{shu2006} that in some situations the ambipolar diffusion may be not fast enough to explain the removal of magnetic fields from accretion disks. This is the case when we claim that the ``reconnection diffusion'' should dominate. At the same time, in cores with low turbulence, the ambipolar diffusion may dominate the reconnection diffusion. The exact range of the parameters for one or the other process to dominate should be defined by future research.

\subsection{Unsolved Problems and Future Studies}

Our paper has a clearly exploratory character. For instance, to simplify the interpretation of our results we studied the concentration of material in the given gravitational potential, ignoring self-gravity of the gas. We plan to study this elsewhere. In addition, our study indicates that the highly magnetized gas in gravitational potential is subject to instabilities (Parker--Rayleigh--Taylor-type) which drive turbulence and induce reconnection diffusion of magnetic field. This is another avenue that we intend to explore.

We report fast magnetic diffusion which happens at the rate of turbulent diffusion, but within the present set of simulations we do not attempt to precisely evaluate the rate. Thus we do not attempt to test, e.g., the predictions in \citet{lazarian2006} of the variations of the turbulent diffusion rate with the fluid magnetization for the passive scalar field. We also observe that while the magnetic field and the passive scalar field diffuse fast, there are differences in their diffusion arising, e.g., from magnetic field being associated with magnetic pressure. We have not attempted to quantify these differences in our work either.

The justification of our results being applicable to molecular clouds is based on the model of fast magnetic reconnection in the partially ionized gas in \citet{lazarian2004}. This model is a natural generalization of the LV99 model, but, unlike the LV99 model, it has not been numerically tested yet. Such testing will be valuable in view of our present study.

\section{Summary}

Motivated by a vital problem of the dynamics of magnetic fields in astrophysical fluids, in particular, by the magnetic flux removal in star formation,  in this paper we have numerically studied the diffusion of magnetic field both in the absence and in the presence of gravitational potential. Recent work on validating the idea of LV99 model of reconnection supports our assertion that our results obtained at moderate resolution represent the dynamics of turbulent magnetic field lines in astrophysics. Our findings obtained on the basis of three-dimensional MHD numerical simulations can be briefly summarized as follows:

1. In the absence of gravitational potential the ``reconnection diffusion'' removes strong anti-correlations of magnetic field and density that we impose at the start of our simulations. The system after several turbulent eddy turnover times relaxes to a state with no clear correlation between magnetic field and density, reminiscent of the observations of the diffuse ISM by \citet{troland1986}.

2. Our simulations that started with a quasi-static
equilibrium in the presence of a gravitational potential, revealed that the turbulent diffusivity induces gas to concentrate at the center of the gravitational potential, while the magnetic field is efficiently pushed to the periphery. Thus the effect of the magnetic flux removal from collapsing clouds and cores, which is usually attributed to ambipolar diffusion effect, can be  successfully accomplished without ambipolar diffusion, but in the presence of turbulence.

3. Our simulations that started in a state of dynamical collapse induced by an external gravitational potential showed that in the absence of turbulence, the flux-to-mass ratio is preserved for the collapsing gas. On the other hand, in the presence of turbulence, fast removal of magnetic field from the center of the gravitational potential occurs. This may  explain the low magnetic flux-to-mass ratio observed in stars compared to the corresponding ratio of the interstellar gas.

4. As an enhanced Ohmic resistivity to remove magnetic flux from cores and accretion disks has been appealed in the literature, e.g., by \citet{shu2006}, we have also compared models with a turbulent fluid and models without turbulence but with substantially enhanced Ohmic diffusivity. We have shown that, in terms of the magnetic flux removal, the reconnection diffusion can mimic the effect of an enhanced Ohmic resistivity.

5. In addition, our results extend earlier findings of \citet{cho2003a} for heat advection by magnetized turbulence. We show that heat advection can be parameterized by the product of the turbulence injection scale and the turbulent velocity for a range of Alfv\'enic and sonic Mach numbers.

\acknowledgments
R.S.L. and E.M.G.D.P. acknowledge partial support from grants of the Brazilian Agencies FAPESP (2006/50654-3 and 2007/04551-0), CAPES (PDEE 3979-08-3), and CNPq. 
A.L. acknowledges the NSF grant AST 0808118 and the NSF-sponsored Center for Magnetic Self-Organization. 
J.C. was supported in part by the National Research Foundation of Korea (NRF) grant funded by the Korea government (MEST) (No. 2009-0077372). Helpful comments by Ellen Zweibel and discussions with Enrique Vazquez-Semadeni and Fabian Heitsch are acknowledged.

\begin{appendix}

\section{Test case: Turbulent advection of heat}

We present here some studies of three-dimensional transport properties of MHD turbulence. Our setup is identical to that adopted in \citet{cho2003a}. Our goal is to provide simulations with higher numerical resolution (the resolution employed in \citealt{cho2003a} is mostly $192^{3}$) and for a larger parameter space.
In the present work, we study how compressibility and magnetization affect the results.
We note that at the time when the results in \citet{cho2003a} were obtained the issues of reconnection in turbulent media were more speculative. Thus, after numerical testing of the model of reconnection in LV99 by \citet{kowal2009}, we feel it is appropriate to revisit that domain.

Besides the ideal MHD set of equations, we also evolve four independent fields of passive scalars $\Phi _{\alpha} (\mathbf{x})$ ($\alpha = 1,2,3,4$) by the continuity equation:
\begin{equation}
\frac{\partial \Phi _{\alpha}}{\partial t}  + \nabla \cdot \left( \Phi _{\alpha} \mathbf{u} \right)  = 0 \text{.}
\label{eqn:continuity_heat}
\end{equation}
These passive scalar fields can trace, for example, the volume concentration of some physical property of the gas, like metalicity or heat\footnote{It may appear that there is some contradiction here between the assumption of an isothermal equation of state (which implies instantaneous heat diffusion) and the adoption of a variable  scalar field to describe heat. However, if we had employed an adiabatic equation of state, the results below would not change sensitively, as long as turbulent transport of heating is concerned (e.g., \citealt{cho2003a}).}, as long as the time-scale associated with the molecular (microscopic) diffusion is larger than the typical dynamical time-scale of the flow. In our simulations this is ensured by fluids being turbulent\footnote{As soon as the molecular diffusivity rate $L^2/\eta$
gets larger than the rates associated with the flow, i.e., $V/L$, the corresponding Reynolds number $Re=VL/\eta$ gets less than unity and turbulence decays at the scale $L$.}.

These passive scalar fields are assumed to have initial spherical symmetry with a Gaussian radial profile:
\begin{equation}
\Phi _{\alpha} (\mathbf{x}, t=0) = \exp \left\lbrace \frac{3}{2} \frac{( \mathbf{x - x_{0}} )^{2}}{\sigma _{0}^{2}} \right\rbrace
\end{equation}
where $\mathbf{x_{0}}$ is the center of the box and, as in \citet{cho2003a}, we choose the initial dispersion $\sigma _{0} = L\sqrt{3} / 16\sqrt{2}$ ($\approx 19$ grid cells, for the resolution employed of $256^{3}$), to ensure that the characteristic width $\sigma _{0}$ is inside the inertial range of the turbulence. This initial distribution of $\Phi _{\alpha}$ is a natural choice to study its diffusion. If we had, for instance, a microscopic diffusion coefficient $\kappa$, the Gaussian shape of the distribution would remain unaltered, with the dispersion increasing linearly in time at a rate $\kappa$.

The initial magnetic field $\mathbf{B} _{0}$ is uniform and parallel to the \textit{x}-direction, and the initial density field is also uniform with $\rho=1$. The passive scalar fields are injected after the turbulence is fully developed. The first scalar-field $\Phi _{1}$ is injected at $t=3$ (approximately $6$ turn-over times), and the other fields ($\alpha = 2, 3, 4$) at each time step after.

Table \ref{tab:diffusion1} shows the set of parameters studied. The magnetic field is given in terms of the initial Alfv\'en speed. Turbulence is injected with the same power $\epsilon=1$ in all the models. These runs cover four combinations of regimes of sonic and Alfv\'enic Mach numbers ($M_{s}$ and $M_{A}$, respectively). The quantities shown in Table \ref{tab:diffusion1} ($M_{s}$, $M_{A}$, $V_{\text{rms}}$, $B_{\text{rms}}$) are averages taken over the time after the injection of the first passive field (these time averages are over the available computed cubic domains  taken at  every $0.25$ time step). The standard deviation of these averages are shown within parentheses. For each data cube, the Mach numbers correspond to the average of the local values computed over the entire box.
The resolution employed in all the simulations is $256^3$.

\begin{table*}[!hbt]
\caption{Simulations of the Turbulent Diffusion of Passive Scalar Fields}
\centering
\begin{tabular}{c c c c c c c c c c}
\hline \hline
Model	&	$c_{s}$	&	$B_{0}$	&	$M_{s}$	&	$M_{A}$	 &	$V_{\text{rms}}$ &	$B_{\text{rms}}$ &	$C_{*}$ &	$C_{*}/V_{\text{rms}}$ &	 Resolution \\
[0.5ex]
\hline
A1	&	$3.0$	&	$0.1$	&	$0.25(0.00)$	&	$2.41(0.06)$	&	$0.75(0.01)$	&	$0.51(0.01)$	&	 $0.4$ &	 $\approx 0.5$ &	$256^{3}$ \\

A2	&	$3.0$	&	$1.0$	&	$0.27(0.01)$	&	$0.78(0.03)$	&	$0.80(0.02)$	&	$1.23(0.01)$	&	 $0.15$ &	 $\approx 0.2$ &	$256^{3}$ \\

A3	&	$0.1$	&	$0.1$	&	$7.53(0.13)$	&	$2.13(0.12)$	&	$0.75(0.01)$	&	$0.38(0.01)$	&	 $0.4$ &	 $\approx 0.5$ &	$256^{3}$ \\

A4	&	$0.1$	&	$1.0$	&	$7.44(0.12)$	&	$0.55(0.01)$	&	$0.74(0.01)$	&	$1.14(0.01)$	&	 $0.3$ &	 $\approx 0.4$ &	$256^{3}$ \\
[1ex]
\hline
\vspace{2mm}
\end{tabular}
\label{tab:diffusion1}
\end{table*}

\subsection{Results}

Compressibility can change the properties of flows substantially. For instance, high Mach number turbulence is known to create regions of enhanced density which will coexist with the expanses of low density. At the same time, one should expect to see similarity between the properties of fluid in low-Mach number flows and incompressible flows.

We first consider a case of heat transport, which can be considered as a test case, as we can compare our results with the earlier heat transport simulations in \citet{cho2003a}. This is the case where the back-reaction can be ignored and we can use a passive scalar diffusion to represent the diffusion of heat.

Let us relate the evolution of the dispersion $\sigma$ with the turbulent diffusion coefficient $\eta _{\text{turb}}$, that is, the coefficient that gives the rate at which the passive scalar field diffuses in scales larger than the turbulent scale. The dispersion $\sigma$ is calculated through the definition: $\sigma^{2} = \frac{\int  (\mathbf{x} - \bar{\mathbf{x}})^2 \Phi(\mathbf{x}) d^{3}x}{ \int \Phi(\mathbf{x}) d^{3}x}$, where $\bar{\mathbf{x}} = \frac{\int  \mathbf{x} \Phi(\mathbf{x}) d^{3}x}{ \int \Phi(\mathbf{x}) d^{3}x}$.  If $\lambda$ is the $rms$ distance between two fluid elements being advected, for $\lambda$ higher than the injection scale of the turbulence $l_{\text{inj}}$, its evolution is related to $\eta _{\text{turb}}$ by $\delta \lambda ^2 \sim \eta _{\text{turb}} \delta t$.

Considering ordinary hydrodynamic turbulence, we can write, for the $rms$ distance $l$ between two fluid elements,
within the inertial range,
\begin{equation}
\delta l^2 \sim v_{l} l \delta t \text{,}
\label{11}
\end{equation}
where $v_{l}$ is the velocity at scale $l$. From the assumption that both relations above should be valid at
the scale $l_{\text{inj}}$, it must be true that $\eta _{\text{turb}} \sim l_{\text{inj}} v_{\text{turb}}$, and we could define a constant $C_{\text{dyn}}$ so that:
\begin{equation}
\eta _{\text{turb}} = C_{\text{dyn}} l_{\text{inj}} v_{\text{turb}} \text{.}
\end{equation}

Now, one wonder if $C_{\text{dyn}}$ is reduced by the magnetic field in magnetized turbulence (due to suppression of turbulent mixture). $C_{\text{dyn}}$ is related to the constant of proportionality in equation (\ref{11}). Again, let $l$ be the distance between two fluid elements initially separated by a distance $l_{0}$ within the inertial range in ordinary hydrodynamic turbulence. Repeating equation (\ref{11}) and using the Kolmogorov's phenomenology,
\begin{equation}
\delta l \sim v_{l} \delta t \sim (\epsilon l)^{1/3} \delta t \text{,}
\end{equation}
where $\epsilon$ is the power of injection (or dissipation) of the kinetic energy. Integrating the last expression,
\begin{equation}
l^{2/3} - l_{0}^{2/3} = C_{R} \epsilon ^{1/3} (t - t_{0}) \text{,}
\label{14}
\end{equation}
where the dimensionless constant $C_{R}$ is the Richardson constant \citep{richardson1926,lesieur1990}. Therefore, $C_{\text{dyn}}$ is related to $C_{R}$.

Using the dispersion $\sigma$ of our experiments as the $rms$ distance $l$ in the equation (A6), and supposing that this equation remains valid in MHD turbulence,
\begin{equation}
\sigma^{2/3} - \sigma _{0}^{2/3} = C_{*} \epsilon ^{1/3} (t - t_{0}) \text{,}
\label{12}
\end{equation}
where $C_{*}$ has not necessarily the same value as that of $C_{R}$.

\begin{figure}[!hbt]
 \begin{center}
 \includegraphics[width=1.0 \columnwidth]{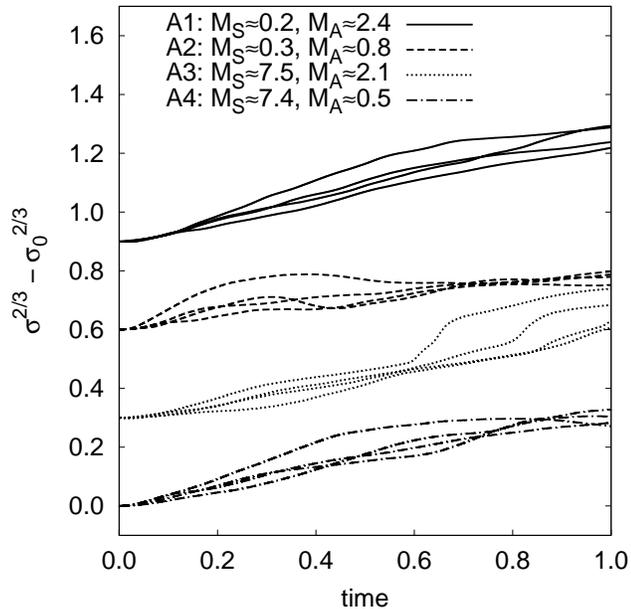}
 \caption{Evolution of $\sigma^{2/3} - \sigma_{0}^{2/3}$ for different models, where $\sigma^{2} = \frac{\int  (\mathbf{x} - \bar{\mathbf{x}})^2 \Phi(\mathbf{x}) d^{3}x}{ \int \Phi(\mathbf{x}) d^{3}x}$ and $\bar{\mathbf{x}} = \frac{\int  \mathbf{x} \Phi(\mathbf{x}) d^{3}x}{ \int \Phi(\mathbf{x}) d^{3}x}$. We have introduced an offset for the lines of each model in order to make the visualization clearer.}
 \label{fig:diffusion2}
 \end{center}
\end{figure}

Figure \ref{fig:diffusion2} shows the evolution of $(\sigma^{2/3} - \sigma_{0}^{2/3})$ for our models. We have made an offset of the values for each model in order to make the visualization easier. Table \ref{tab:diffusion1} lists the approximated values of $C_{*}$ for the models, obtained from a linear fit of the data.

All in all, our results presented in Figure \ref{fig:diffusion2} are consistent with the scalings of diffusion shown in \citet{cho2003a}. This indicates that the relation (\ref{12}) continues approximately valid in the MHD case. The slopes do not seem to be very sensitive to $M_{s}$, but we can note a small difference for the trans-Alfv\'enic and super-Alfv\'enic cases, the constant $C_{*}$ being a little higher in the super-Alfv\'enic cases. This finding is consistent with the prediction of changes of the diffusion efficiencies of \citet{lazarian2006}. There it is shown that for higher magnetization, corresponding to an Alfv\'en Mach number $M_A=V/V_A<1$, where $V$ is the injection velocity and $V_A$ is the Alfv\'enic velocity, the turbulence at the large scales is weak, which produces weak mixing. In our parameterization this corresponds to the decrease of $C_{*}$, which is the case of A2 simulation. Interestingly enough, in the case of $M_A<1$ and $M_s\gg 1$, which is the case of A4 simulation, we see the increase of $C_{*}$. The case of high Mach number turbulence is not covered by \citet{lazarian2006}, but we expect that the concentration of gas in dense regions plays an important role for the effect. In regions of gas concentration the Alfv\'enic velocity drops and the corresponding $M_A$ increases (see \citealt{burkhart2009}). As a result, the mixing should be less constrained by the magnetic field as we see in the A4 model. Comparing cases A1 and A3 we observe that for super-Alfv\'enic turbulence and different Mach number $M_s$ the diffusivity is rather similar.

Note that \citet{cho2003a} obtained the values of $C_{*}/V_{\text{rms}}=0.4$ ($V_{A}=0.1$ and $M_{s}=2.3$), $0.3$ ($V_{A}=1$ and $M_{s}=2.3$), $0.4$ ($V_{A}=1$ and $M_{s}=0.3$), and $0.5$ (hydrodynamic case and $M_{s}=0.3$), which are consistent with our findings. In \citet{cho2003a}, $V_{\text{rms}} \approx 0.8$ in all the cases (to make a comparison we had to convert the time-scale to the present code units, i.e. 1 time step here = 2$\pi$ time steps there).
We conclude that, for a wide range of magnetizations and Mach numbers, turbulence is efficient in inducing turbulent diffusivity of heat.

Our results show that the turbulent diffusivity of passive scalar fields $\eta _{\text{turb}}$ is well described by $\eta _{\text{turb}} = C_{*} v_{\text{turb}} l_{\text{inj}}$, similar to the hydrodynamic case, and that the coefficient $C_{*}$ is not very sensitive to the strength of the magnetic field. Therefore, the magnetic field does not impose a strong suppression of the turbulent diffusion perpendicular to it as far as heat transfer is concerned.

As already pointed in \citet{cho2003a} (see also \citealt{lazarian2006}), the effectiveness of the turbulent diffusion has important consequences on the thermal diffusion in the ISM and intracluster medium (ICM). The nearly isotropic turbulent diffusivity can be 1 order of magnitude higher in the gas of central regions of galaxy clusters, like the Hydra A cluster, compared to the \textit{Spitzer} value, based on the kinetic theory. In the ISM, in mixing layers, where turbulence is very strong, the turbulent diffusivity can be 2 orders of magnitude higher than the laminar values (see \citealt{esquivel2006}). Our results extend the range of the applicability of the turbulent heat advection model of \citet{cho2003a}.

\section{Diffusion of passive scalar in the absence of gravity}

In our simulations described in Section 4 (see Table \ref{tab:heitsch2}) we also keep track of passive scalar field representative either of heat or passive impurity.

Figure \ref{fig:heitsch8} (left panels) shows the time evolution of $\left\langle B_{z} \right\rangle _{0.25} / \left\langle \rho \right\rangle _{0.25} - \bar{B_{z}} / \bar{\rho} $ and $\left\langle \Phi \right\rangle _{0.25} / \left\langle \rho \right\rangle _{0.25} - \bar{\Phi} / \bar{\rho}$. Both quantities have a similar behavior, and all models seem to achieve the characteristic average values ($\bar{B_{z}}/\bar{\rho}$ and $\bar{\Phi}/\bar{\rho}$) roughly at the same time.

%
Right plot of Figure \ref{fig:heitsch8} is an analog of Figure \ref{fig:heitsch8B}. A comparison between these figures reinforces that the quantity $\left\langle B_{z} \right\rangle _{z} / \left\langle \rho \right\rangle _{z}$  evolves similar to $\left\langle \Phi \right\rangle _{z} / \left\langle \rho \right\rangle _{z}$. How can we understand this behavior?

Using the expression $\frac{\partial {\bf B}}{\partial t}=\nabla\times({\bf v}\times {\bf B})$ one gets $\frac{\partial {\bf B}}{\partial t}= ({\bf B}\cdot\nabla){\bf v}-({\bf v}\cdot\nabla){\bf B}$, when $\nabla\cdot{\bf B}=0$ is accounted for. Using the continuity equation in the form
$\nabla\cdot{\bf v}= -\frac{1}{\rho}\frac{\partial \rho}{\partial t}-\frac{{\bf v}}{\rho}\cdot\nabla\rho$ one can get
\begin{equation}
\left(\frac{\partial}{\partial t} + {\bf v}\cdot\nabla\right) \frac{\bf B}{\rho}=\frac{\bf B}{\rho} \cdot \nabla {\bf v}
\end{equation}
which coincides with Equation \ref{eqn:continuity_heat} if $\Phi_{\alpha}$ is substituted by ${\bf B} / \rho$.

Figure \ref{fig:heitsch11b} (left panels) shows the distribution of $\left\langle \rho \right\rangle _{z}$ versus $\left\langle \Phi \right\rangle _{z}$ for the model B2 ($B_{0}=1.0$, see Table \ref{tab:heitsch2}) at different time steps ($t=0$, $10$). The initial anti-correlation gives place to a tight correlation between the passive scalar and density. Right side of Figure \ref{fig:heitsch11b} shows the evolution of $\left\langle \delta \Phi, \delta \rho \right\rangle$ (defined by equation \ref{eqn:correlation}, where $B$ must be substituited by $\Phi$).

\section{Further comparison of magnetic diffusivity and resistivity effects in the presence of gravity}

\begin{figure*}[!hbt]
 \begin{center}
 \includegraphics[width=0.95 \textwidth]{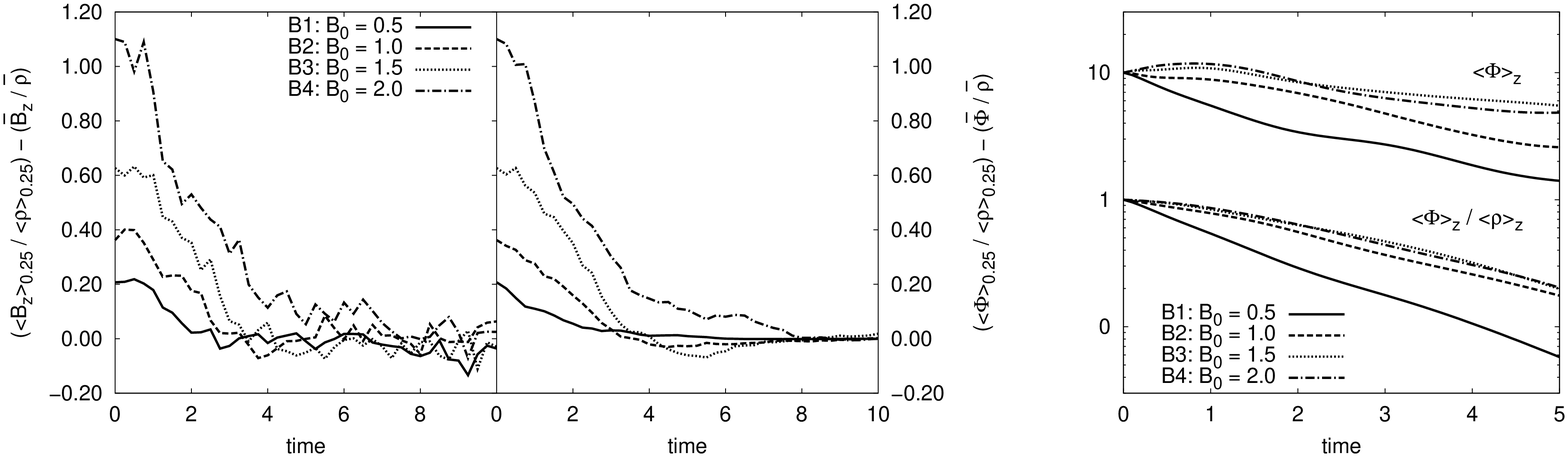}
 \caption{\textit{Left:} evolution of the ratio of the averaged magnetic field over the averaged density (more left) and of the ratio of the averaged passive scalar over the averaged density (more right) within a distance $R=0.25L$ from the central \textit{z}-axis. The values have been subtracted from their characteristic values $\bar{B} / \bar{\rho}$ in the box. \textit{Right:} evolution of the $rms$ amplitude of the Fourier modes $(k_{x},k_{y})=(\pm 1,\pm 1)$ of $\left\langle \Phi \right\rangle _{z}$ (upper curves) and $\left\langle \Phi \right\rangle _{z} / \left\langle \rho \right\rangle _{z}$ (lower curves). The curves for $\Phi$ were multiplied by a factor of $10$. All the curves were smoothed to make the visualization clearer. \label{fig:heitsch8}}
\vspace{5mm}
 \includegraphics[width=0.95 \textwidth]{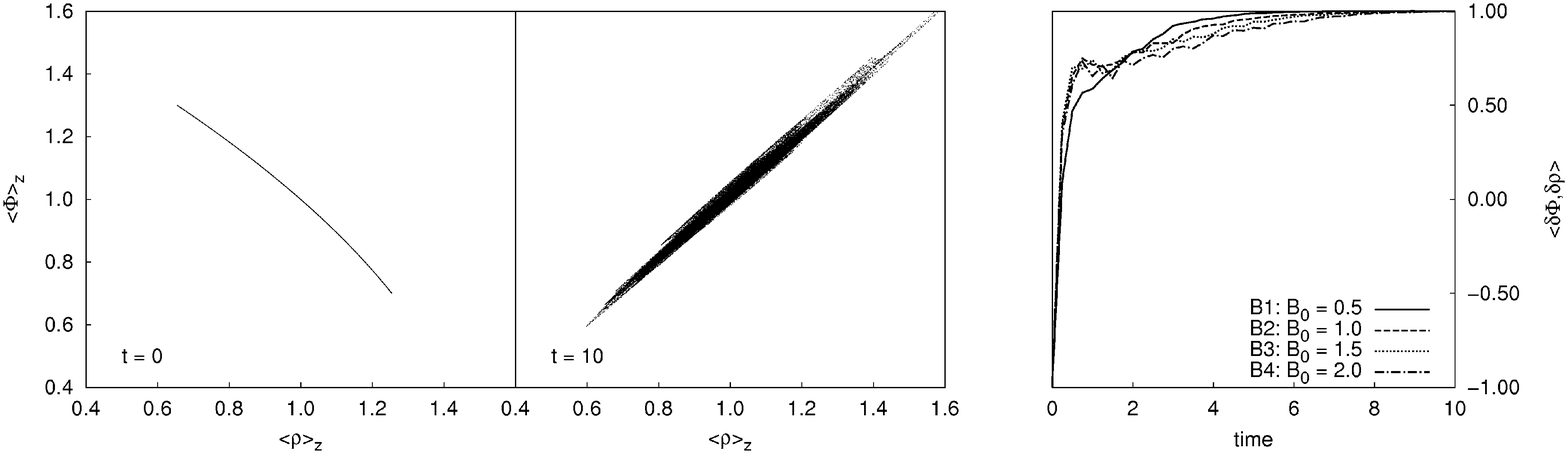}
 \caption{\textit{Left}: distribution of $\left\langle \rho \right\rangle _{z}$ vs. $\left\langle \Phi \right\rangle _{z}$ for model B2 (see Table \ref{tab:heitsch2}), at $t=0$ (most left) and $t=10$ (most right). \textit{Right}: correlation between fluctuations of the passive scalar field ($\delta \Phi$) and density ($\delta \rho$). \label{fig:heitsch11b}}
\vspace{5mm}
 \includegraphics[width=0.95 \textwidth]{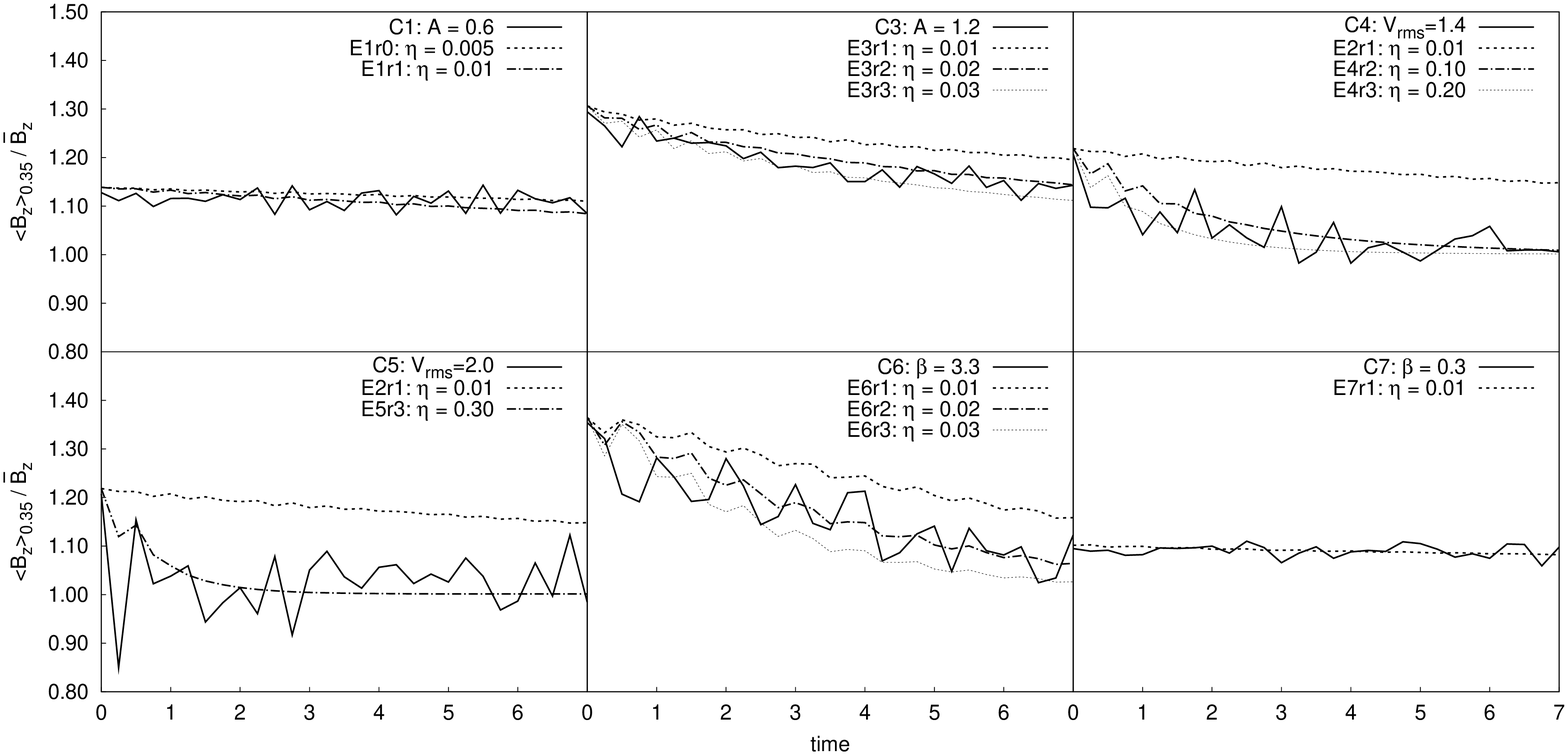}
 \caption{Comparison of the time evolution of $\left\langle B_{z} \right\rangle _{0.35}$ between models C1, C3, C4, C5, C6, and C7 (see Table \ref{tab:cylind}) and resistive models without turbulence (see Table \ref{tab:cylind3}). \label{fig:ohmic2}}
 \end{center}
\end{figure*}

Figure \ref{fig:ohmic2} shows the comparison of the evolution of $\left\langle B_{z} \right\rangle _{R=0.35L}$ for models C1, C3, C4, C5, C6, and C7 of Table \ref{tab:cylind} with similar resistive models without turbulence of Table \ref{tab:cylind3}, with different values of Ohmic diffusivity $\eta_{\text{Ohm}}$. From these comparisons (and others where $\left\langle B_{z} \right\rangle _{R}$ is measured for different values of $R$, like in Figure \ref{fig:ohmic1}), we have estimated the values of $\eta_{\text{turb}}$ shown in Table \ref{tab:cylind}.

%
\end{appendix}

\end{document}